\documentclass[a4paper,twocolumn,showpacs,prb,superscriptaddress,floatfix
]{revtex4}
\usepackage{graphicx,color}
\usepackage[normalem]{ulem}
\usepackage{amssymb}
\usepackage{amsmath,amsfonts,latexsym}
\usepackage{array,tabularx}
\usepackage{dcolumn}               

\newcommand{\vect}[1]{{\mathbf #1}}
\newcommand{\Frac}[2]{\displaystyle\frac{#1}{#2}}

\newcommand{\epstil}{\tilde{\epsilon}}
\newcommand{\nodagger}{\vphantom{\dagger}}
\renewcommand{\Im}{\mathrm{Im}}
\renewcommand{\Re}{\mathrm{Re}}

\begin{document}

\title{Mean field theory and fluctuation spectrum of a pumped,
  decaying Bose-Fermi system across the quantum condensation
  transition} 
\author{M.~H.~Szyma{\'n}ska}
\affiliation{Clarendon Laboratory, Department of Physics, University of Oxford,
             Parks Road, Oxford, OX1 3PU, UK}
\author{J.~Keeling\footnote{Present Address: Cavendish Laboratory, 
    University of Cambridge, Madingley Road, Cambridge CB3 0HE, UK}}
\affiliation{Department of Physics, Massachusetts Institute of
  Technology, 77 Mass. Ave., Cambridge, MA 02139, USA}
\author{P.~B.~Littlewood}
\affiliation{Cavendish Laboratory, University of Cambridge,
             Madingley Road, Cambridge CB3 0HE, UK}
\begin{abstract}
\end{abstract}
\begin{abstract}
  We study the mean-field theory, and the properties of fluctuations,
  in an out of equilibrium Bose-Fermi system, across the transition to
  a quantum condensed phase.  The system is driven out of equilibrium
  by coupling to multiple baths, which are not in equilibrium with
  each other, and thus drive a flux of particles through the system.
  We derive the self-consistency condition for an uniform condensed
  steady state.  This condition can be compared both to the laser rate
  equation and to the Gross-Pitaevskii equation of an equilibrium
  condensate.  We study fluctuations about the steady state, and
  discuss how the multiple baths interact to set the system's
  distribution function.  In the condensed system, there is a soft
  phase (Bogoliubov, Goldstone) mode, diffusive at small momenta due
  to the presence of pump and decay, and we discuss how one may
  determine the field-field correlation functions properly including
  such soft phase modes.  In the infinite system, the correlation
  functions differ both from the laser and from an equilibrium
  condensate; we discuss how in a finite system, the laser limit may
  be recovered.
\end{abstract}
\maketitle

\section{Introduction}
\label{Int}

In the last decade there have been enormous advances in the experimental
realisation and theoretical understanding of the phenomenon of quantum
condensation, i.e macroscopic occupation of a single quantum mode, in different
physical conditions.
The phenomena ranges from Bose-Einstein Condensation (BEC) of structureless
bosons to the BCS-type collective state of fermions and has been studied in
several physical systems such as degenerate atomic gases and
superconductors~\cite{SnokeBook}.
Further, recent experimental advances in manipulation of atomic Fermi
gases have led to realisation of the BCS-BEC crossover regime
\cite{Regal04,Zweierlein04} and low-dimensional atomic condensates
have also been explored\cite{Gorlitz01,Paredes04,Kinoshita04,Stock05}.
From the early days of experimental investigation of BEC there have
been enormous efforts in order to realise quantum condensation
in the solid state \cite{SnokeBook}.
For this, the currently promising candidates are excitons in coupled
quantum wells~\cite{Butov02a,Butov02b,Snoke02}, microcavity
polaritons~\cite{Dang98,Deng,Richard}, quantum Hall
bilayers~\cite{QH}, and Josephson junction arrays in microwave
cavities~\cite{JJ}. Although all these systems potentially may
condense at temperatures orders of magnitude higher than those for
dilute atomic gases, it has proven to be much more difficult to
realise BEC in the solid state than in atomic traps. Recently a
comprehensive set of experiments \cite{Kasprzak06} reports polariton
condensation in CdTe based microcavities but still the level of
control in the study of the condensed states in solid-state is far
from the finesse achieved in atomic vapours.

In these various candidates for condensation, one should
distinguish different classes of systems.  
Equilibrium superconductors are special in that the decay of pairs is
disallowed.
In equilibrium particle-hole condensates, such as quantum Hall
bilayers or charge density waves, particle-hole mixing (tunnelling in
bilayers) leads to a gapped spectrum; however the gap may be very
small.
Non-equilibrium particle-hole condensates in the solid state are, to a
much greater extent than atomic gases, subject to dephasing and
decay.
It is not usually possible to isolate the condensate
from the environment:
lattice phonons, impurities and imperfections of the crystal structure lead to
dephasing, and due to poor trapping, particles escape, requiring external
pumping to sustain a steady-state.
The dephasing and decay processes are often faster than thermalisation, putting
the system out of thermal equilibrium. 
The decay, and consequent lack of equilibrium have for a long time
presented the major experimental obstacle in the realisation of
solid-state condensation in otherwise appropriate conditions.
Even if one can accelerate thermalisation\cite{Kasprzak06,Hui06},
comparing the decay rates to other energy scales, one may see that
decay, and the consequent flux of particles through the system,
remains a more important effect in solid state than in atomic gases.

Thus, a significant presence of dissipation and decay also poses
fundamental questions about the robustness of a condensate, for
example: whether a steady-state condensate is possible with incoherent
pumping and decay, and if so, how does it differ from thermal
equilibrium, and from a laser\cite{keldysh_letter}.
Quantum condensation in dissipative systems also provides a connection to
other phenomena of collective behaviour in the presence of dissipation such as
pattern formation\cite{CrossHohenberg,Haken:RMP}, particularly in
lasers\cite{Haken70,Staliunas,Denz03},
and also recently in a system related
to that studied here, the coherently pumped polariton optical parametric
oscillator\cite{Wouters05,Wouters06}.
Other recent examples of phase transitions and coherence in driven
systems include quantum criticality in magnetic systems in the
presence of currents\cite{Mitra,Hogan}, and transport through a Kondo
dot coupled to multiple dots\cite{Paaske}.
The relation between lasing and BEC is particularly relevant for
polariton BEC, where the experimental distinction between the two is
not straightforward~\cite{Szymanska}.

Microcavity polaritons in particular, being made from fermionic
particles and photons, have several special features and so provide an
excellent laboratory to  study condensation in dissipative
environment.
Due to the large wave-length of their photonic component and non-linearities
associated with underlying fermionic structure the physics exits the regime of
weakly interacting bosons at even modest density \cite{Keeling}. 
Putting aside a few subtleties characteristic only for polaritons one
can say that with increasing density the quantum condensation
transition moves from BEC (fluctuation dominated) to something like
the BCS (mean-field, interaction dominated) collective state
\cite{Keeling}, analogous to the BCS-BEC crossover in atomic Fermi
gases near Feshbach resonance \cite{Regal04,Zweierlein04}.
This allows one to explore the influence of non-equilibrium and
dissipation not only on the usual BEC but also on more exotic forms of
quantum condensation.
A further complication is that
polaritons in planar microcavities are two-dimensional(2D) particles
and so in an infinite equilibrium system, although there is a
Berezhinskii-Kosterlitz-Thouless (BKT) transition to a superfluid phase,
below the transition long-wavelength fluctuations destroy the
off-diagonal long range order and result in algebraic decay of phase
coherence. 
Dissipation changes the structure of collective modes and influences
the spatial and temporal coherence in 2D quasicondensates, changing
the power-law controlling the decay of phase correlations
\cite{keldysh_letter}.
Finally, microcavity polaritons can also be trapped either in
stress-induced harmonic potentials\cite{Snoke06,Daif06,Baas06} or in
natural traps provided by microcavity disorder which reduce the
influence of long-wavelength fluctuations and may allow the existence
of a true condensate and phase coherence over the whole system size
\cite{Kasprzak06}.
How this confinement, when combined with pumping and decay, modifies the
properties of coherence in such systems is an interesting
question\cite{savaona06}, which has not yet been fully addressed.

The last issue is particularly relevant for the deeper understanding
of the differences and connections between a polariton
condensate and the laser.  Apart from the obvious difference; the
laser being a collective coherent state of massless non-interacting
photons while the condensate consists of massive and interacting
bosons (in polariton condensation both massive photons and strongly
coupled excitons are coherent), there are more subtle differences
connected with fluctuations and so expected differences in the decay
of correlations \cite{keldysh_letter}. Lasing is normally considered
in systems with a well-defined single or a few mode structure and so
the phase fluctuations which control the laser linewidth are those of
a phase diffusion of a single mode\cite{Haken:Laser}.  In contrast,
condensation is usually studied in systems where there is a continuum
of single particle modes, and thus collective excitations involve
coherent interaction of these different modes which affects the decay
of coherence and the line-shape of the emission \cite{keldysh_letter}.
While lasing in systems with transverse freedom has been investigated
for its pattern forming properties\cite{Denz03}, there remain many
open questions concerning the decay of correlations and the crossover
from a small system with few spatial modes to the infinite and
many-mode limit.

Although semiconductor microcavities in strong coupling provide a
natural system to explore such phenomena, all these issues are by no
means restricted to polariton condensation. With recent advances in
manipulating dilute atomic gases similar conditions can be engineered,
an immediate example is that of an atom laser in which a continuous
leakage of atoms from atomic BEC takes place. To our knowledge the
description of the output from an atom laser has been to date largely
analogous to that of the photon laser\cite{Holland96} and the
influence of the continuum of modes connected with atomic BEC on the
coherence properties of the atom laser has not been addressed.

In a previous paper\cite{keldysh_letter} we addressed some of
these issues.
We used a model Bose-Fermi system
coupled to independent baths, not in thermal or
chemical equilibria with each other, providing incoherent pumping and decay.
We show that steady-state spontaneous condensation can occur in such
systems, and can be distinct from lasing:
The condensate can exist at low densities, far from the inversion
required for lasing.
We also found that the collective modes are qualitatively
altered by the presence of pumping and decay:
The low energy phase mode (Goldstone, Bogoliubov mode) becomes diffusive at
small momenta.
By considering the effect of phase fluctuations, we 
described the decay of correlations, which at large times and
distances differs both from that for a thermal equilibrium condensate
and from a laser.

In this manuscript, apart from providing technical details of the
method we address several new aspects of quantum condensation in
dissipative environment.  In particular we study the influence of the
exciton density of states, and the temperature of the pumping bath on
the non-equilibrium phase diagram.
We also analyse how the non-thermal occupation of photon
states is controlled by competition between the pumping and decay baths,
and how this occupation deviates from that in
thermal equilibrium. 
 We do not a priori assume that the system is close to
equilibrium, and so the system's distribution function may
be of any form.
Finally we provide a full account of how to determine field-field
correlation functions in the condensed state, where phase fluctuations
may be large, and so expansion to second order is insufficient.
These field-field correlation functions describe the decay of
correlations at large times and distances, and their Fourier transform
gives the line-shape of a non-equilibrium condensate.
This is an important extension to the non-equilibrium path integral
techniques which to our knowledge has not been done before.
In the final section of this paper we study how dissipation influences
spatial and temporal coherence in a finite size condensate and show
how the linewidth of emission from polariton or atom condensate should
be determined taking proper account of the spatial fluctuations.
We further emphasise the fundamental difference between emission from
a polariton condensate or an atom laser and that from the photon
laser.
 
The paper is organised as follows:
The model for the system, and for the reservoirs to which it is
coupled is introduced in Sec.~\ref{Model}, then in
Sec.~\ref{sec:path-integr-form} we show how to integrate out first the
reservoirs, and then the fermionic fields to give an effective action
in terms of the photon field. 
We then study this effective action in the saddle-point approximation
in Sec.~\ref{sec:saddle-point-mean}. 
In Sec.~\ref{sec:second-order-fluct}, by discussing fluctuations about
the saddle point we consider the stability of the saddle-point
solutions, and show how the instability of the normal state, and the
photon distribution functions, compare to an equilibrium treatment.
Having identified the stable and unstable saddle-point solutions,
Sec.~\ref{sec:numer-analys-mean} then presents numerical results for
the critical conditions at which steady-state, non-equilibrium
condensation occurs.
The effects of fluctuations on correlation functions in the condensed
case are studied again in Sec.~\ref{sec:fluct-cond-state}, where
care is taken to correctly describe phase fluctuations in the broken
symmetry system.
Section~\ref{sec:finite-size-effects} then studies how finite size
modifies correlation functions, and the relation between the previous
results and laser theory.
Finally, section~\ref{sec:conclusions} summarises our results.

\section{Model}
\label{Model}

Our Hamiltonian is
\begin{equation}
\hat{H}=\hat{H}_{\rm{sys}}+\hat{H}_{\rm {sys,bath}}+\hat{H}_{\rm{bath}},
\label{Hmain}
\end{equation} 
where,
\begin{multline}
  \hat{H}_{\rm{sys}} 
  = 
  \sum_{\alpha} \epsilon_{\alpha}
  \left(
    b_{\alpha}^\dag b_{\alpha}^{} - a_{\alpha}^{\dag}a_{\alpha}^{}
  \right)
  + 
  \sum_{\vect{p}} 
  \omega^{}_{\vect{p}} \psi_{\vect{p}}^\dag \psi_{\vect{p}}^{}
\\ 
  + 
  \Frac{1}{\sqrt{L^2}} 
  \sum_{\alpha} \sum_{\vect{p}} \left(
    g^{}_{\alpha, \vect{p}} \psi_{\vect{p}}^{} b_{\alpha}^\dag a_{\alpha}^{} +
    \text{H.c.}
  \right) 
\label{Hsys}
\end{multline}
describes two fermionic species $b_{\alpha}$ and $a_{\alpha}$,
interacting with bosonic modes $\psi_{\vect{p}}$ normalised in a
2D box of area $L^2$, with $L\to\infty$.
Condensed solutions of Eq.~(\ref{Hsys}) have been
studied in the context of atomic Fermi gases
\cite{Holland01,Timmermanns01,Ohashi02} and microcavity
polaritons~\cite{Eastham,Keeling,Marchetti}.
In this work we focus on microcavity polaritons, and so this
  model describes the interaction between disorder-localised
  excitons which are dipole coupled to cavity photon modes
  $\psi_{\vect{p}}$, with low $\vect{p}$ dispersion,
  $\omega_{\vect{p}} \simeq \omega_0 + \vect{p^2}/2 m_{\text{ph}}$,
  where $m_{\text{ph}} = (\hbar/c)(2\pi/w)$ is the photon mass in a
  2D microcavity of width $w$.
  The disorder localised excitons are described here as in previous
  works~\cite{Eastham,Keeling,Marchetti} by hard-core bosons; i.e. the
  Coulomb interaction between excitons is described by exclusion,
  preventing multiple occupation of a single disorder-localised
  state $\alpha$.
  This hard core boson is represented by a two-level system,
  described here as two fermionic levels,
  $b^{\dagger}_{\alpha}, a^{}_{\alpha}$.
  Thus, the combination $b^{\dagger}_{\alpha} a^{}_{\alpha}$ creates
  an exciton in the localised state with energy $\epsilon_{\alpha}$.
  This energy $\epsilon_{\alpha}$ includes the Coulomb binding within
  an exciton state.
  In such a description, it is important not to confuse the fermion
  states (representing a hard-core bound exciton) with the underlying
  conduction and valence band states (see e.g.\
  Refs.~\onlinecite{SST,Marchetti} for further discussion of this
  point).
  In order that these fermionic levels describe a two-level system, it
  is necessary that the constraint $b^{\dagger}_{\alpha} b^{}_{\alpha}
  + a^{\dagger}_{\alpha} a^{}_{\alpha} = 1$ is satsified; i.e. that
  exactly one of the two levels is occupied.
  In thermal equilibrium, this constraint can be exactly imposed by a
  shift of Matsubara frequencies\cite{Popov88}; and in that case it
  can be easily seen that the difference between imposing the single
  occupancy constraint exactly and imposing it on average leads only
  to a factor of 2 in the definition of temperature.
  Out of thermal equilibrium, no simple shift to the Matsubara
  frequencies is possible, although an extension to the
  non-equilibrium case has been proposed\cite{Kiselev00}. 
  For simplicity, in this work, we will impose the single occupancy
  constraint on average,  as discussed below when introducing
  the occupation functions of the bath.

\begin{figure}[htpb]
  \centering
  \includegraphics[width=0.98\columnwidth]{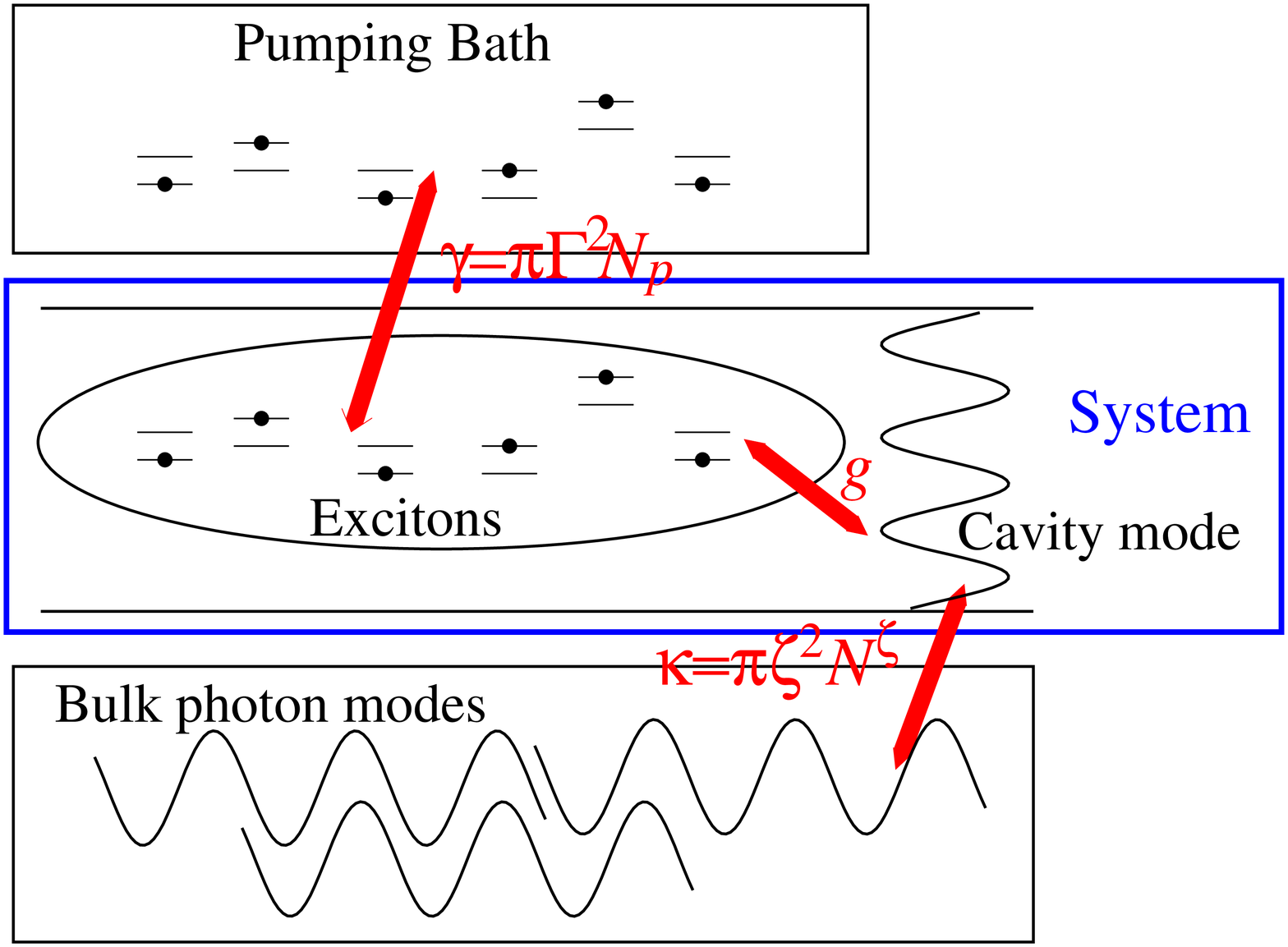}
  \caption{(Color online) Schematic diagram illustrating parts of the
    Fermi-Bose system, and its coupling to baths.  The parts in the
    box labelled system are described by Eq.~(\ref{Hsys}), while the
    effective couplings to the baths, described by
    Eq.~(\ref{Hsysbath}), lead to effective pump and decay rates
    $\gamma$ and $\kappa$ as discussed later.}
\label{fig:schematic}
\end{figure}

Because of the imperfect reflectivity of the cavity mirrors, photons
escape, so the system must be pumped (excitons injected) to sustain a
steady-state.
As illustrated schematically in Fig.~\ref{fig:schematic}, the
imperfect reflectivity of the mirrors is represented by
coupling to the continuum of bulk photon modes.
Incoherent fermionic pumping is described by coupling
to a pumping bath ---  which is represented
mathematically as two separate fermionic baths, coupled
to the two fermionic modes.
Thus, the coupling of the system to these pumping and decay
baths is written as:
\begin{multline} 
  \hat{H}_{\rm{sys,bath}}
  = 
  \sum_{\alpha,k}
  \Gamma^{a}_{\alpha,k} (a^\dagger_\alpha A_{k}^{} +
  \text{H.c.} ) +
  \Gamma^{b}_{\alpha,k}
  (b^\dagger_\alpha B_{k}^{} + \text{H.c.}) 
  \\ 
  +
  \sum_{\vect{p},k}
  \zeta^{}_{\vect{p},k}(\psi_{\vect{p}}^\dagger \Psi_{k}^{} +
  \text{H.c.}),
\label{Hsysbath}
\end{multline} 
Here $A_{k}^{}, B_{k}^{}$ are fermionic annihilation
operators for the pump baths, while $\Psi_{k}^{}$ are bosonic annihilation
operators for photon modes outside the cavity.  
The Hamiltonian corresponding to the evolution of these baths 
is given by:
\begin{equation}
\hspace{-0.005\textwidth}
  \label{Hbath}
  \hat{H}_{\rm{bath}}  = 
  \sum_{k}\omega^{\Gamma^a}_{k}
  A_{k}^{\dagger}A_{k}^{} +
  \sum_{k}\omega^{\Gamma^b}_{k}
  B_{k}^{\dagger}B_{k}^{} +
  \sum_{k}\omega^\zeta_{k}\Psi_{k}^\dagger \Psi_{k}^{}.
\end{equation}

The pumping bath, if thermalised at some finite non-zero temperature,
acts both as a source of particles, and also tries to drive the
polariton distribution function towards a thermal distribution in
equilibrium with the bath.
In some physical systems, one might also consider a bath which purely
provides a thermalisation mechanism, such as phonons, which
redistribute energy, but do not change particle number.
We do not explicitly consider such a bath.
However, in the example of microcavity polaritons, our model may still
capture much of the important behaviour, for the following reason.
One may consider the low energy polaritons as being pumped by a
reservoir of higher energy excitons.
These excitons are formed by the binding of the electrons and holes injected
by the pump laser, and subsequent relaxation by phonon emission, and are
thus partially thermalised.
By regarding our pumping bath as describing a partially thermalised
exciton reservoir, our model, being interacting, could thus describe
the thermalisation of low energy polaritons pumped by such a
reservoir.

Although in the absence of other processes, the excitons would
thermalise to the pumping bath, they are also strongly coupled to
photons, which in turn couple to a second environment of the bulk
photon modes outside the cavity.  The strongly coupled exciton-photon
system would be therefore influenced by two independent environments
which are not in thermal or chemical equilibrium with each other. Even
in the steady-state, if the rates of dissipation to the environment
are larger than the polariton-polariton interactions, the system would
remain out of thermal equilibrium. In addition, even if the
thermalisation via polariton-polariton interaction is fast, so the
system distribution function would be close to thermal, particles are
continuously added and removed from the system. We show that this
particle ``current'' has dramatic consequences on the properties of
such a condensate even if it remains close to equilibrium.

We would like to stress that there are two distinct issues, both of which we
intend to address.  The first is that of non-equilibrium distribution
functions, in systems where the internal thermalisation rate is slower than
the pumping and decay rates --- i.e. when the coupling to the external baths
is strong, and the baths are not in equilibrium with each other.  The second
issue is the presence of particle ``current'' in strongly dissipative systems
--- even if internal thermalisation rates are large, this current may be
important if the pumping, decay and thermalisation rates are large compared to
other energy scales.

In the next section we will introduce the path integral formalism which will
allow us to treat the nonequilibrium conditions.  Our approach will then be to
assume that the pumping and decay baths are much larger than
the system, and so the populations in the baths are fixed.  This will enable
us to describe the properties of the system as influenced by its coupling to
the baths.  These influences modify
both the system's spectrum and the population of this spectrum.  We will look
for steady states of the system in the presence of pumping and decay, and
study the excitation spectra around these steady states.

\section{Path Integral Formulation}
\label{sec:path-integr-form}

In order to study the system away from thermal equilibrium, we proceed
using the path-integral formulation of non-equilibrium Keldysh field
theory, as described in detail in Ref.~\onlinecite{Kamenev}.
Following the prescription there, we write the quantum partition
function as a coherent state path integral over bosonic and fermionic
fields defined on a closed-time-path contour, $\mathcal{C}$.
 Arranging the fermionic fields into a Nambu vector
$\bar{\phi}=( \bar{b}, \bar{a})$ and $\phi= (b,a)^{\rm T}$,
loosely referred to as ``particle/hole'' space, the partition
function can be formally written as:
\begin{multline*} 
{\cal Z}= {\cal N} \int \prod_{\vect{p}} D[\bar{\psi}_{\vect{p}},
  \psi_{\vect{p}}] \prod_\alpha D[\bar{\phi}_\alpha,\phi_\alpha] \\ \times
  \prod_k D[\bar{A}_k, A_k, \bar{B}_{k}, B_{k},\bar{\Psi}_{k}, \Psi_{k} ]
  e^{iS}, 
\end{multline*}
where ${\cal N}$ represents a constant of normalisation and the total
action can be separated into constituent components
$S=S_\phi+S_\psi+S_{{\rm bath},\phi}+S_{{\rm bath},\psi}$.
The part:
\begin{equation*} 
\hspace{-0.005\textwidth}
  S_\phi = \int\limits_{\cal C} dt \sum_{\alpha, \vect{p}}
\bar{\phi}_\alpha \left[i\partial_t-\epsilon_\alpha \sigma_3^{
 }-g_{\alpha,\vect{p}}\bar{\psi}_\vect{p}
\sigma_-^{ } - g_{\alpha,\vect{p}}\psi_\vect{p}
\sigma_+^{ } \right]\phi_\alpha,
\end{equation*}
describes the free exciton evolution together with the dipole interaction
between excitons and photons.
Due to the Nambu formalism, the term in brackets is a matrix,
and has been decomposed in terms of the Pauli matrices $\sigma_i$ 
operating in the particle-hole $(b,a)$ space (with $\sigma_0=1$).
The time derivative is taken along the Keldysh contour ${\cal C}$.
Similarly,
\begin{equation*}
S_\psi = 
\int\limits_{\cal C} dt\;
\sum_{\vect{p}}
\bar{\psi}_\vect{p}\left(i\partial_t
-\omega_{\vect{p}}\right)\psi_{\vect{p}}
\end{equation*}
describes the free photon dynamics. 
The excitonic environment and the interactions between excitons and
their environment is given by
\begin{eqnarray*} 
S_{{\rm bath},\phi}  = \int\limits_{\cal C} dt \sum_{\alpha,k} [ 
\bar{A}_{k}\left(i\partial_t 
-\omega^{\Gamma^a}_{k}\right)A_{k}+
\bar{B}_{k}\left(i\partial_t 
-\omega^{\Gamma^b}_{k}\right)B_{k} & & \\
 -\Gamma^{b}_{\alpha,k}
(\bar{b}_\alpha B_{k} + \bar{B}_{k} b_\alpha) - \Gamma^{a}_{\alpha,k}
(\bar{a}_\alpha A_{k} + \bar{A}_{k} a_\alpha) ],  
\end{eqnarray*}
while the photonic environment is given by
\begin{multline*}
S_{{\rm bath},\psi}=\int\limits_{\cal C} dt \sum_{\vect{p},k} 
[   
\bar{\Psi}_k\left(i\partial_t -
\omega^{\zeta}_k\right)\Psi_{k}  - \\ \zeta_{\vect{p},k}
\left(\bar{\psi}_\vect{p}\Psi_k+\bar{\Psi}_k\psi_\vect{p}\right) ]. 
\end{multline*}

As described in Ref.~\onlinecite{Kamenev}, the standard procedure is to
replace the fields on the closed-time-path contour by a doublet of fields
$\psi=(\psi_f, \psi_b)$ on the forward and backward branches.
This then leads to four Green's functions:
forward $iG^{<}(t,t^{\prime})= \langle \psi^{ }_f (t)
\psi^{\dagger}_b(t^{\prime}) \rangle$, backward
$iG^{>}(t,t^{\prime})=\langle \psi^{ }_b (t)
\psi^{\dagger}_f(t^{\prime}) \rangle$, time-ordered
$iG^{T}(t,t^{\prime})=\langle \psi^{ }_f
(t)\psi^{\dagger}_f(t^{\prime}) \rangle $, and anti-time-ordered
$iG^{\tilde{T}}(t,t^{\prime})= \langle \psi^{ }_b (t)
\psi^{\dagger}_b(t^{\prime}) \rangle$. 
In the homogeneous steady-state these are functions of $r-r^{\prime}$
and $t-t^{\prime}$ alone and when transformed into $\vect{p}$ and
$\omega$ space, in the case of photon fields, the functions $iG^{<}$
and $iG^{>}$ give the luminescence and absorption spectra
respectively.
Again following Ref.~\onlinecite{Kamenev}, as these four Green's functions
are not independent, one proceeds by making a rotation to {\it
  classical} $\phi_{cl}=(\phi_f+\phi_b)/\sqrt{2}$ and {\it quantum}
$\phi_{q}=(\phi_f-\phi_b)/\sqrt{2}$ components. 
All fields are from now vectors in Keldysh space, i.e 
$\psi=(\psi_{cl},\psi_{q})$, and we define an
additional matrix in Keldysh $(cl,q)$ space:
\begin{eqnarray*}
\psi^M={\frac{1}{\sqrt{2}}}\left(
\begin{array}{cc}
\psi_{cl} & \psi_{q} \\ \psi_{q} & \psi_{cl}
\end{array}
\right)=\psi_{cl}\sigma_0^K+\psi_q\sigma_1^K,
\end{eqnarray*}
(where $\sigma_i^K$ are Pauli matrices in Keldysh space).
One may then write the action as:
\begin{multline*} 
S_\phi=\int_{-\infty}^\infty dt \sum_{\alpha,\vect{p}}
\\
\bar{\phi}_\alpha [i\partial_t-\epsilon_\alpha \sigma_3^{ }-
\frac{g_{\alpha,\vect{p}}}{\sqrt{2}}\bar{\psi}_\vect{p}^M\sigma_-^{ }
- 
\frac{g_{\alpha,\vect{p}}}{\sqrt{2}}\psi_\vect{p}^M\sigma_+^{ }
]\sigma_1^{\textsc{k}}\phi_\alpha,
\end{multline*}
\begin{multline*}
\begin{aligned}
S_\psi=\int_{-\infty}^\infty dt\; \sum_\vect{p}
\bar{\psi}_\vect{p}\left(i\partial_t
-\omega_\vect{p}\right)\sigma_1^{\textsc{k}}\psi_\vect{p},
\end{aligned}
\end{multline*}
\begin{multline*} 
S_{{\rm bath}\phi}= \int_{-\infty}^\infty dt \sum_{\alpha,k} [
  -\Gamma^{b}_{\alpha,k} (\bar{b}_\alpha \sigma_1^{\textsc{k}} B_{k} +
  \bar{B}_{k} \sigma_1^{\textsc{k}} b_\alpha)  \\
  -\Gamma^{a}_{\alpha,k} (\bar{a}_\alpha \sigma_1^{\textsc{k}} A_{k} +
  \bar{A}_{k} \sigma_1^{\textsc{k}} a_\alpha) \\ +
  \bar{B}_{k}\left(i\partial_t
  -\omega^{\Gamma^b}_{k}\right)\sigma_1^{\textsc{k}} B_{k}  +
  \bar{A}_{k}\left(i\partial_t -\omega^{\Gamma^a}_{k}\right)
  \sigma_1^{\textsc{k}} A_{k} ],
\end{multline*}
\begin{multline*}
S_{{\rm bath}\psi}=\int_{-\infty}^\infty dt \sum_{\vect{p},k} [ -
\zeta_{\vect{p},k} \left(\bar{\psi}_\vect{p}
\sigma_1^{\textsc{k}}\Psi_k+\bar{\Psi}_k\sigma_1^{\textsc{k}}\psi_\vect{p}
\right) \\ + \bar{\Psi}_k\left(i\partial_t -
\omega^{\zeta}_k\right)\sigma_1^{\textsc{k}}\Psi_{k}].
\end{multline*}  

\subsection{Treatment of environment}
\label{sec:treatm-envir}

As we are interested in the properties of the system, rather than the
properties of the baths,  we next integrate over the bath fields, to leave
an effective action expressed only in terms of the fields describing
the system.   
If the  baths are much larger than the system, then
their behaviour is not affected by the interaction with the system.
One may then evaluate correlation functions of bath operators as for
free bosons and free fermions; these correlators in turn depend on the
distribution function of the baths, i.e. the population of the bath
modes.
The effects of the environment then enter as self energies for the
system fields,  which modify both the spectrum and its occupation.
This procedure is described in Ref.~\onlinecite{Kamenev};  we summarise
the results here both to show how it applies to our system, and also
as our notation differs slightly from Ref.~\onlinecite{Kamenev}
%
%
For the decay (photon) bath one has:
\begin{multline*}
S_{{\rm bath}\psi}= -\int \int_{-\infty}^\infty dt dt^{\prime}
\sum_{\vect{p},\vect{p^{\prime}}} 
\\
\bar{\psi}_\vect{p}(t)\sigma_1^{\textsc{k}}
\sum_k \zeta_{\vect{p},k} \zeta_{\vect{p^{\prime}},k}\left[(i\partial_t -
\omega^{\zeta}_k)\sigma_1^{\textsc{k}}\right]^{-1}\sigma_1^{\textsc{k}}
\psi_{\vect{p^{\prime}}}(t^{\prime}),
\end{multline*}
In Keldysh space the Green's function for a free boson has the
following form
\begin{displaymath}   
  \left[(i\partial_t -
    \omega^{\zeta}_k)\sigma_1^{\textsc{k}} 
  \right]^{-1} 
  = 
  \left(
    \begin{array}{cc}\hat{D}_k^K(t-t^{\prime})
      &\hat{D}_k^R(t-t^{\prime}) \\ \hat{D}_k^A(t-t^{\prime})&0 \end{array}
  \right),
\end{displaymath}
where (after the Fourier transform with respect to $t-t^{\prime}$)
the retarded, advanced and Keldysh Green's functions are respectively
\begin{align*}
\hat{D}_k^{R/A}(\omega) = & \frac{1}{\omega-\omega^{\zeta}_k\pm i0},\\
\hat{D}_k^K(\omega) = & (-2\pi i) (2n_B(\omega^{\zeta}_k)+1)
\delta(\omega-\omega^{\zeta}_k). 
\end{align*}
If the bath distributions are thermal, then $n_B$ would be the Bose
occupation functions, however one can also consider arbitrary
function for $n_B$.

Let us now make a number of restrictions on the photon bath,
to simplify the analysis.
Firstly, we will assume that $S_{\text{bath}\psi}$ does not contain
terms off-diagonal in $\vect{p} \vect{p^{\prime}}$.
This means that each confined photon mode $\vect{p}$ couples to a
separate set of bulk photon modes, i.e. that $\zeta_{\vect{p},k}
\zeta_{\vect{p^{\prime}},k} = 0$ unless $\vect{p}=\vect{p^{\prime}}$.
Physically, this can be interpreted as conservation of in-plane
momentum in the coupling of two-dimensional microcavity photon modes to 
bulk modes.
Next, we restrict to the case that all $\vect{p}$ photonic modes
couple to their environments with the same strength i.e
$\zeta_{\vect{p},k}=\zeta_k$.
Then, if the bath frequencies $\omega^{\zeta}_k$ form a dense
spectrum, and the coupling constants $\zeta_k$ are smooth functions of
the frequencies, we may replace the sum over bath modes by an
integral,
\begin{displaymath}
  \sum_{k}\zeta_{k}^2 \to \int d
  \omega^{\zeta}\zeta(\omega^\zeta)^2 N^\zeta(\omega^\zeta),
\end{displaymath}
where we have introduced $N^\zeta(\omega^\zeta)$ as the bath's density
of states.
After integrating over $\omega^\zeta$ we obtain
\footnote{
In taking the Fourier transform ${\cal F}$, we have used
${\cal F}(\bar{\psi}(t-t^{\prime}))=\bar{\psi}(-\omega)$ and 
${\cal F}(\psi(t-t^{\prime}))=\psi(\omega)$.}
\begin{displaymath}
S_{{\rm bath}\psi}=-\int_{-\infty}^\infty d\omega  \sum_{\vect{p}}
\bar{\psi}_\vect{p}(\omega)\left(\begin{array}{cc} 
0 & d^A \\
 d^R & d^K
\end{array} \right)_{(-\omega)}
\psi_\vect{p}(-\omega).
\end{displaymath}
By writing $d^{R,A}(\omega)= R(\omega)  \mp i\kappa(\omega)$ we may
split the bath self energy into an imaginary part, describing
broadening
\begin{displaymath}
  \kappa(\omega)=\pi\zeta^2(\omega)N^{\zeta}(\omega),
\end{displaymath}
and a real energy shift,
\begin{displaymath}
R(\omega)=\int d \omega^{\zeta}
\frac{\zeta^2(\omega^{\zeta})N(\omega^{\zeta})}{\omega-\omega^{\zeta}}.
\end{displaymath}
In terms of these, the Keldysh component becomes:
$d^K(\omega)= -i2\kappa(\omega)(2n_B(\omega)+1)$.

Although the formalism allows one to consider any density of states,
and coupling strength as a function of frequency, one possible choice
is a Markovian (or Ohmic) bath --- i.e.  a white noise environment
--- where the density of states for the bath and the coupling constant
of the system to the bath are frequency independent, and so
$\zeta^2(\omega^{\zeta}) N^{\zeta}(\omega^{\zeta})=\zeta^2N^{\zeta}$.
For this case the real energy shift $R(\omega)$ is zero while
$\kappa(\omega)=\kappa$.  
In this work, we will consider this Markovian limit, but
due to the bath's occupation function, the Keldysh component
will remain frequency dependent.
Combining the free photon action with the effective action for the
photon decay, using $F_{\psi}(\omega)=2n_B(\omega)+1$, one has:
\begin{multline*}
S_{\psi}+S_{{\rm bath}\psi}=
\int_{-\infty}^\infty d\omega \sum_{\vect{p}}\\
\bar{\psi}_\vect{p}(\omega) 
\left(
  \begin{array}{cc}
    0 & -\omega -\omega_\vect{p} - i \kappa  \\
    -\omega -\omega_\vect{p} +i \kappa & 2i\kappa F_{\psi}(-\omega)
  \end{array} 
\right) 
\psi_\vect{p}(-\omega).
\end{multline*}

One can follow a similar procedure for the baths connected with the
pumping process.
\begin{multline*}
S_{{\rm bath}\phi}= -\int \int_{-\infty}^\infty dt dt^{\prime}
\sum_{\alpha,\alpha^{\prime}} 
\\
\bar{b}_{\alpha}(t)\sigma_1^{\textsc{k}}\sum_k
\Gamma^b_{\alpha,k}\Gamma^b_{\alpha^{\prime},k}  \left[(i\partial_t -
\omega^{\Gamma^b}_k)\sigma_1^{\textsc{k}} \right]^{-1}
\sigma_1^{\textsc{k}} b_{\alpha^{\prime}}(t^{\prime}) 
\\
+
\bar{a}_\alpha(t)\sigma_1^{\textsc{k}}\sum_k
\Gamma^a_{\alpha,k}\Gamma^a_{\alpha^{\prime},k} \left[(i\partial_t -
\omega^{\Gamma^a}_k)\sigma_1^{\textsc{k}}
\right]^{-1}\sigma_1^{\textsc{k}} a_{\alpha^{\prime}}(t^{\prime}).
\end{multline*}
The Green's function for a free fermion is
\begin{displaymath}   
  \left[(i\partial_t -
    \omega^{\Gamma}_k)\sigma_1^{\textsc{k}} 
  \right]^{-1} 
  = 
  \left(\begin{array}{cc}\hat{P}_k^K(t-t^{\prime})
      &\hat{P}_k^R(t-t^{\prime}) \\ \hat{P}_k^A(t-t^{\prime})&0 \end{array}
  \right),
\end{displaymath}
where in frequency space 
\begin{align*}
\hat{P}_k^{R/A}(\nu) = & \frac{1}{\nu-\omega^{\Gamma}_k\pm i0},\\
\hat{P}_k^K(\nu) = & (-2\pi i) (1-2n_F(\omega^{\Gamma}_k))
\delta(\nu-\omega^{\Gamma}_k).
\end{align*}
In the same way as above, $n_F$ would be the Fermi occupation function
for a thermal distribution.

For compact notation, we will define additional matrices in (b,a)
space as $\sigma_{\uparrow} = \left(\begin{array}{cc} 1 & 0 \\ 0 & 0
\end{array}\right)$ and $\sigma_{\downarrow} = \left(\begin{array}{cc} 
0 & 0 \\ 0 & 1 \end{array}\right)$, and so:
\begin{align*}
  S_{{\rm bath}\phi}
  &= 
  \sum_{\alpha,\alpha^{\prime}}
  \int_{-\infty}^\infty d \nu 
  \bar{\phi}_\alpha(\nu) 
  \Sigma^{\Gamma}_{\alpha,\alpha^{\prime}}(\nu)
  \phi_{\alpha^{\prime}}(-\nu),
  \\
%
  \Sigma^{\Gamma}_{\alpha,\alpha^{\prime}}
  &=
  \left(
    \begin{array}{cc}
      0 & p_{b,\alpha,\alpha^{\prime}}^A\sigma_{\uparrow}^{ } +
      p_{a,\alpha,\alpha^{\prime}}^A\sigma_{\downarrow}^{ } \\
      p_{b,\alpha,\alpha^{\prime}}^R\sigma_{\uparrow}^{ }
      +p_{a,\alpha,\alpha^{\prime}}^R\sigma_{\downarrow}^{ } &
      p_{b,\alpha,\alpha^{\prime}}^K \sigma_{\uparrow}^{ }
      +p_{a,\alpha,\alpha^{\prime}}^K \sigma_{\downarrow}^{ }
    \end{array} \right)
\end{align*}
where
$p_{b/a,\alpha,\alpha^{\prime}}^{R/A/K}=\sum_k
\Gamma^{b/a}_{\alpha,k}\Gamma^{b/a}_{\alpha^{\prime},k} P_k^{R/A/K}$ 
with the fermionic propagators $P_k^{R/A/K}$ as defined earlier.
Now we make similar restrictions as for the photonic environment: we
consider all excitons coupled equally strongly to the environment (the
coupling constants of the system to the bath is $\alpha$ independent)
and take the Markovian limit.
Without much loss of generality we can further assume that the
coupling strength of the two fermionic species to their pumping
baths are the same, after all of which
$\Gamma_{\alpha,k}^{b/a}=\Gamma$.
As in the bosonic case, in the Markovian limit, the real self
energy shift vanishes, and imaginary part takes the form:
\begin{displaymath}
  \gamma=\pi \Gamma^2 N_p,
\end{displaymath}
with $N_p$ being the bath's density of states.
Of course, due to the distribution function of the bath, despite the
Markovian limit, the effective pumping rate of a given exciton state
will depend on its energy.
The final form is then:
\begin{align*}
  \Sigma^{\Gamma}_{\alpha,\alpha^{\prime}}(\nu)
  &=
  \left(
    \begin{array}{cc}
      0 & - i \gamma \sigma_{0}^{ } \\ 
      i\gamma \sigma_{0}^{ } & 
      2 i \gamma ( F_b(-\nu)
      \sigma_{\uparrow}^{ } + F_a(-\nu)
      \sigma_{\downarrow}^{ })  
    \end{array} 
  \right) 
\end{align*}
where $F_b(\nu)=1-2n^b_F(\nu)$ and $F_a(\nu)=1-2n^a_F(\nu)$ are the
fermion distribution functions.

\begin{figure}[htbp]
  \centering
  \includegraphics[width=0.85\columnwidth,clip]{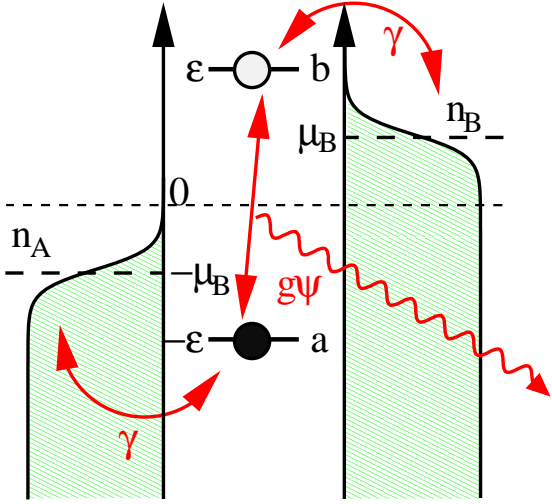}
  \caption{(Color online) Schematic diagram illustrating occupation of
    A and B baths, and their coupling to the two fermionic levels.  In
    the case shown, the bath chemical potential is below the energy
    level $\epsilon$, so the pumping cannot lead to inversion.  The
    flow of energy is from the pumping bath, through the fermionic
    levels of an exciton, to the photons, and energy is then
    lost into the photon bath.}
  \label{fig:occupations}
\end{figure}

Any functions for the bath distribution $n^b_F$ and $n^a_F$ can be
considered within this formalism.  One physical choice, as illustrated
in Fig.~\ref{fig:occupations}, can be pumping of
quantum-well excitons by contact with some
thermal reservoir with a chemical potential
$\mu_B$, i.e
\begin{equation}
F_b(\nu)=\mathrm{tanh}\frac{\beta}{2}(\nu-\mu_B) \ \ \ \
F_a(\nu)=\mathrm{tanh}\frac{\beta}{2}(\nu+\mu_B),
\label{FaFb}
\end{equation}
where $\beta=1/kT$. 
Note that, as discussed earlier, these bath distributions have
  been chosen so that on average, $\langle b^{\dagger} b^{} +
  a^{\dagger} a^{} \rangle = 1$;  i.e. the single occupancy constraint
  for these fermionic states to represent two-level systems 
  is obeyed on average.
In the absence of any other processes, contact between the excitons
and the pumping reservoir would control the population of
excitons, and so
\begin{displaymath}
  \langle b^{\dagger}b-a^{\dagger}a \rangle
  =
  n^b_F(\epsilon)-n^a_F(-\epsilon)
  =
  -\mathrm{tanh}\frac{\beta}{2}(\epsilon-\mu_B).
\end{displaymath}
Thus, by pumping with a thermalised source of electrons, one will find
a thermalised distribution of excitons.

Before proceeding further, let us examine what the form of the self
energy due to the bath tells us about the relation between
thermalisation and dephasing.
In principle one could consider a non-Markovian environment where for
some range of frequencies one has $\gamma(\omega)=0$,
$\kappa(\omega)=0$.
Then, for that range, there would be no damping, but also $p^K(\omega)
= 0$, $d^K(\omega) = 0$, so the system distribution in such a range
would not be influenced by the bath --- i.e. no thermalisation.
Thus for a full thermalisation of all relevant modes of the system,
one needs a non-zero coupling to the low frequency modes of the
environment, which will at the same time introduce dephasing.

\subsection{Integration over fermionic fields}
\label{sec:integr-over-ferm}

After eliminating the bath's degrees of freedom the full action $S$ becomes
\begin{multline*}
  S=
  \int_{-\infty}^\infty dt dt^{\prime} 
  \left[
    \sum_{\alpha,\alpha^{\prime}}
    \bar{\phi}_\alpha(t) G_{\alpha,\alpha^{\prime}}^{-1}(t,t^{\prime})
    \phi_{\alpha^{\prime}}(t^{\prime})
  \right.
  \\
  +
  \left.
    \sum_\vect{p} 
    \bar{\psi}_\vect{p}(t) 
    \left(
      \begin{array}{cc}
        0 & i\partial_t -\omega_\vect{p} - i\kappa \\
        i\partial_t -\omega_\vect{p} +i\kappa & 2i\kappa F_{\Psi}(t-t^{\prime})
      \end{array} 
    \right) 
    \psi_\vect{p}(t^{\prime})
  \right],
\end{multline*}
where $G$ is the exciton Green's function.
Introducing the abbreviations
\begin{math}
  \lambda_{cl}=\sum_{\vect{p}}
  \frac{g_{\vect{p},\alpha}}{\sqrt{2}}\psi_{\vect{p},cl}
\end{math}
and
\begin{math}
  \lambda_{q}=\sum_{\vect{p}}
  \frac{g_{\vect{p},\alpha}}{\sqrt{2}}\psi_{\vect{p},q}  
\end{math}
we may write $G$ as:
\begin{widetext}
  \begin{multline}
    G_{\alpha,\alpha^{\prime}}^{-1}(t,t^{\prime}) = \\
    \left(
      \begin{array}{cc}
        (\lambda_q(t)\sigma_{+}^{ } +
        \bar{\lambda}_q(t)\sigma_{-}^{ })\delta_{\alpha,\alpha^{\prime}} &
        (\partial_t \sigma_{0}^{ } -\epsilon_{\alpha} \sigma_{3}^{ }
        -\lambda_{cl}(t)\sigma_{+}^{ } -
        \bar{\lambda}_{cl}(t)\sigma_{-}^{ } )
        \delta_{\alpha,\alpha^{\prime}}- i \gamma 
        \sigma_{0}^{ } \\
        (i\partial_t \sigma_{0}^{ } -\epsilon_{\alpha} \sigma_{3}^{ }
        -\lambda_{cl}(t)\sigma_{+}^{ }
        - \bar{\lambda}_{cl}(t)\sigma_{-}^{ } )
        \delta_{\alpha,\alpha^{\prime}} + i\gamma \sigma_{0}^{ } &
        -(\lambda_q(t)\sigma_{+}^{ } +
        \bar{\lambda}_q(t)\sigma_{-}^{
        })\delta_{\alpha,\alpha^{\prime}} + 2 i \gamma ( 
        F_b \sigma_{\uparrow}^{ } + F_a \sigma_{\downarrow}^{ })
      \end{array} \right), 
    \label{Gtotal}
  \end{multline} 
\end{widetext}
Note that $F_b(t-t^{\prime})$, $F_a(t-t^{\prime})$ as well as
$i\partial_t$ are time non-local.
It is now explicit how the competition between the two environments
works. 
The photon environment, with the distribution function $F_\Psi$ of
modes outside of the cavity, affects the free photon evolution.
Similarly, the fermionic environment, with the bath distributions
$F_b, F_a$, enters the exciton Green's function, now also modified by
the presence of cavity photons.
The spectrum of this coupled system will combine both the strong
coupling between excitons and photons as well as the dissipation to
the environment.
The occupation of these modes will be a non-trivial combination of the
distributions of the baths as well as the exciton-photon interaction.

The action S is quadratic in fermionic fields and therefore it is
possible to integrate over the fermionic degrees of freedom $\phi$
and, in the same spirit as in previous studies of the equilibrium
properties of this model~\cite{Eastham,Szymanska,Keeling,Marchetti},
obtain the total effective action for the photon field alone
\begin{multline}
\hspace{-0.02\textwidth}S = 
 -i\sum_{\alpha}{ \rm Tr ln}
G_{\alpha,\alpha}^{-1}
+
\int_{-\infty}^\infty dt dt^{\prime} \sum_{\vect{p}}
\\
\bar{\psi}_\vect{p}(t) 
\left(
\begin{array}{cc}
0 & i\partial_t -\omega_\vect{p} - i\kappa \\
 i\partial_t -\omega_\vect{p} +i\kappa & 2i\kappa F_{\Psi}(t-t^{\prime})
\end{array} \right)   
 \psi_\vect{p}(t^{\prime}) 
\label{totalS}
\end{multline}
Other than those approximations explicitly discussed in the text,
this expression is exact; i.e.\ it makes no assumption about
what form $\psi_{\vect{p}}(t)$ takes.
Note, however, that due to the non-linear term ${ \rm Tr ln}
G_{\alpha,\alpha}^{-1}$ the action is highly complex and contains all
powers of $\psi_{cl}$ and $\psi_q$. Therefore some expansion scheme
needs to be performed.

\section{Saddle-Point (Mean-Field) Analysis}
\label{sec:saddle-point-mean}

In order to determine the state of the pumped, decaying, strongly
coupled system, we will follow a standard method for path integrals,
and first find the saddle point solution.
The saddle-point equations for the action (\ref{totalS}) have
the following form
\begin{multline*}
\hspace{-0.02\textwidth}
  \frac{\delta S}{\delta \bar{\psi}_{\vect{p},q}}
  = \int dt^{\prime}\biggl\{
    \left[
      (i\partial_t-\omega_\vect{p})\delta(t-t^{\prime})-d^R(t-t^{\prime})
    \right]
    \psi_{\vect{p},cl}(t^{\prime})
  \biggr.
  \\
  \biggl.
    - d^K(t-t^{\prime}) \psi_{\vect{p},q}(t^{\prime})
  \biggr\}
  -
  \sum_\alpha \frac{g_{\vect{p},\alpha}}{\sqrt{2}}(-i){ \rm Tr}
  (G_{\alpha,\alpha} \sigma_-^{ } \sigma_0^{\textsc{k}}) = 0,
\end{multline*}
\begin{multline*}
\hspace{-0.02\textwidth}
  \frac{\delta S}{\delta \bar{\psi}_{\vect{p},cl}} 
  = \int dt^{\prime} 
  \left[
    (i\partial_t-\omega_\vect{p})\delta(t-t^{\prime})-d^A(t-t^{\prime})
  \right]
  \psi_{\vect{p},q}(t^{\prime}) \\
  -\sum_\alpha\frac{g_{\vect{p},\alpha}}{\sqrt{2}}(-i){ \rm Tr}
  (G_{\alpha,\alpha} \sigma_-^{ } \sigma_1^{\textsc{k}}) = 0.
\end{multline*}
It can be seen that the second equation is satisfied by
$\psi_{\vect{p},q}=0$ (classical saddle-point).
Putting $\psi_{\vect{p},q}=0$ into (\ref{Gtotal}) gives the usual
structure for the mean-field exciton Green's function, which ensures
causality \cite{Kamenev}:
\begin{displaymath}
  G^{-1}=
  \left(
    \begin{array}{cc}
      0 & [G^A]^{-1} \\ 
      \left[G^R\right]^{-1} & \left[G^{-1}\right]^K
    \end{array} 
  \right)
\end{displaymath}
and so 
\begin{math}
  G=
  \left(
    \begin{array}{cc}
      G^K & G^R \\ G^A & 0
    \end{array} 
  \right),
\end{math}
where $G^K=-G^R[G^{-1}]^KG^A$.
It is now clear why the Keldysh rotation discussed earlier, i.e.
working in terms of $(cl,q)$ components rather than $(f,b)$,  is
more convenient.
By reducing the number of dependent functions, both the
Green's function and inverse Green's function contain a zero
block, and so become easier to invert.
With this structure
${ \rm Tr} (G \sigma_-^{ } \sigma_0^{\textsc{k}})={\rm Tr}(G^K
\sigma_-^{ })$
and
$
{ \rm Tr} (G \sigma_-^{ } \sigma_1^{\textsc{k}})={\rm Tr}((G^R+G^A)
\sigma_-^{ }) =0
$
since $G^R(t,t)+G^A(t,t)=0$. 
Thus we are left with only the first of the saddle point equations,
which now becomes:
\begin{multline*}
\int dt^{\prime}\left[(i\partial_t-\omega_\vect{p})\delta_{t-t^{\prime}}-d^R(t-t^{\prime})
\right]\psi_{\vect{p},cl}(t^{\prime}) = \\ \sum_\alpha
\frac{g_{\vect{p},\alpha}}{\sqrt{2}}(-i){ \rm Tr}
(G_{\alpha,\alpha}^K \sigma_-^{ }).
\end{multline*}
Since we consider an infinite homogeneous system (no trap) we expect a
uniform saddle point. We therefore consider the solutions to be of the
form $\psi_\vect{p}=\psi \delta(p)$.
It is difficult to invert $[G^{-1}]^{R,A}$ matrix for an arbitrary
time dependence of the $\psi$ fields.
We are, however, interested in the non-equilibrium
steady-state so we take the only time dependence of the photon field
$\psi$ to be oscillation at a single frequency. Therefore we propose the
following ansatz
\begin{equation}
\psi(t)=\psi e^{-i\mu_S t}.
\label{ansatz}
\end{equation}
Substituting this ansatz in the action of Eq.~(\ref{totalS}) will lead
to explicit time dependence within the exciton inverse Green's
function.
This time dependence can be removed straightforwardly by implementing
an appropriate gauge transformation, described by the following matrix
in particle-hole space:
\begin{displaymath}
  U=\left(
    \begin{array}{cc}
      e^{i\frac{\mu_S}{2}t} & 0 \\ 0 & e^{-i\frac{\mu_S}{2}t}
    \end{array} 
  \right)
\end{displaymath}
The trace is invariant under unitary transformations, ${\rm Tr}\ln
G^{-1} = {\rm Tr}\ln U G^{-1} U^{\dagger}$, so the effects of such
time-dependence appear in only two places: Firstly in the time
derivative terms, which lead to the energy shifts
\begin{math}
  \omega_\vect{p} \to \tilde{\omega}_\vect{p} = \omega_\vect{p}-\mu_s
  \ \ {\rm and} \ \
  \epsilon_\alpha \to \tilde{\epsilon}_\alpha  = \epsilon_\alpha-\mu_s/2,
\end{math}
and secondly in a gauge transformation of the bath functions
\begin{math}
  d^{R/A/K}(\omega) \to d^{R/A/K}(\omega+\mu_S),
\end{math}
\begin{math}
  p_b^{R/A/K}(\nu) \to p_b^{R/A/K}(\nu+\frac{\mu_S}{2}) \ \ {\rm
    and} \ \ p_a^{R/A/K}(\nu) \to
  p_a^{R/A/K}(\nu-\frac{\mu_S}{2}).
\end{math}
In practice, the latter substitutions mean replacing
$\mu_B \to \tilde{\mu}_B = \mu_B - \mu_S/2$ in $F_a,F_b$.

With $\psi_q=0$ and the above time-dependence described by the gauge
transformation, the matrix $G^{-1}$ can now be easily inverted and the
final form for the mean-field exciton Green's functions are
\begin{equation}
G^{R/A}(\nu)= \frac{(\nu \pm i\gamma)\sigma_{0}^{ }+
\tilde{\epsilon}_{\alpha} \sigma_{3}^{ } + g\psi_f
\sigma_{+}^{ } +g \bar{\psi}_f \sigma_{-}^{ }}
{\nu^2-E_\alpha^2 \pm 2i\gamma\nu - \gamma^2
},
\label{GMF1}
\end{equation}
and
\begin{equation}
\hspace{-0.01\textwidth}
G^K_{bb/aa}(\nu)\!=\! -2i\gamma
\frac{F_{b/a}(\nu)[(\nu \!\pm\! \tilde{\epsilon}_\alpha)^2\!+\!\gamma^2] +
F_{a/b}(\nu) g^2|\psi_f|^2
}{[(\nu-E_\alpha)^2+\gamma^2][(\nu+E_\alpha)^2+\gamma^2]},
\label{GMF3}
\end{equation}
\begin{multline}
  G^K_{ba}(\nu)= -(G^{K}_{ab}(\nu))^\ast= -2i\gamma g\psi_f \times \\
  \frac{(F_a(\nu)+F_b(\nu))\nu
    +(F_b(\nu)-F_a(\nu))(\tilde{\epsilon}_\alpha+i\gamma)}
  {[(\nu-E_\alpha)^2+\gamma^2][(\nu+E_\alpha)^2+\gamma^2]},
  \label{GMF5}
\end{multline}
where $a$, $b$ defines the particle-hole space as follows
\begin{displaymath}
G=
\left(
\begin{array}{cc}
G_{bb} & G_{ba} \\ G_{ab}& G_{aa}
\end{array} 
\right),
\end{displaymath}
and $E_\alpha=\sqrt{\tilde{\epsilon}_\alpha^2+g^2|\psi_f|^2}$. Note
that only the site-index diagonal, i.e. ($\alpha$,$\alpha$), component
appears in the gap equation and so we have omitted the site index in G
for brevity. Also since at the saddle point $\psi_q=0$ then
$\psi_f=\frac{\psi_{cl}+\psi_{q}}{\sqrt{2}}=
\frac{\psi_{cl}}{\sqrt{2}}$.
In this work we consider $g_{\vect{p},\alpha}=g$. The influence of the
distribution of the oscillator strength has been addressed in
Ref.~\onlinecite{Marchetti}.
The mean-field exciton Green's functions physically correspond to
excitons strongly renormalised by the presence of the mean-field
photon field, and damped by the coupling to the environment. The
Keldysh Green's function, which contains the distribution of excitons,
depends on the distributions of the pumping bath.  In general
\begin{displaymath}
G^K=G^RF-FG^A,
\end{displaymath}
where $F$ has a meaning of the quasi-particle distribution function.
We can determine the mean-field distribution function for excitons in
a self-consistent photon field from Eqs.~(\ref{GMF1})--(\ref{GMF5}):
\begin{displaymath}
  F_{bb/aa}(\nu)=
  \frac{F_a(\nu)+F_b(\nu)}{2} 
  \pm 
  \frac{(F_b(\nu)-F_a(\nu))(\tilde{\epsilon}_{\alpha}^2+\gamma^2)}{2(E_{\alpha}^2+\gamma^2)}, 
\end{displaymath}
\begin{displaymath}
  F_{ba}(\nu)=(F_{ab})^\ast(\nu) = 
  \frac{g\psi_f(F_b(\nu)-F_a(\nu))(\tilde{\epsilon}_{\alpha}+i\gamma)}{2(E_{\alpha}^2+\gamma^2)},  
\end{displaymath}
where $F_a$ and $F_b$ are the bath's distributions given by Eq.~(\ref{FaFb}),
with $\mu_B\to\tilde{\mu}_B$.
Note that, since this is a mean-field approximation, only coherent
photons enter in this distribution.
Thus, in the uncondensed case where $\psi=0$ the exciton distributions
reduce to $F_{bb/aa}=F_{b/a}$ and $F_{ab}=0$: i.e. in the absence of
coherent photons, the mean field approximation neglects the effect of
photons on the exciton distribution.
The distribution of the photonic environment will however enter the
distribution of fluctuations about the mean-field, as will be
discussed in Sec.~\ref{sec:norm-state-excit}.

With the ansatz (\ref{ansatz}) the saddle-point equation becomes
\begin{equation}
(\omega_0 -\mu_S -i \kappa) \psi_{f} =  \sum_\alpha
\frac{g}{2}(-i){ \rm Tr}
(G_{ba}^K).
\label{Gap}
\end{equation}
As in equilibrium this is a self-consistent equation for the order
parameter (condensate). 
With changing density the type of transition moves from interaction
dominated BCS-like mean-field regime to a fluctuation dominated BEC
limit\cite{Keeling} (strictly speaking BKT in 2D) . 
So the above equation is analogous to the gap equation in the theory
of BCS-BEC crossover. 
Here it relates the coherent photon field with the exciton Green's
function strongly modified by the presence of such a coherent field.
Physically it means that the coherent field is generated by a coherent
polarisation in the exciton system which in turn is generated by the
presence of the coherent field. 
Thus equation (\ref{Gap}) can be viewed as a non-equilibrium
generalisation of the gap equation. 
One difference with respect to equilibrium is that the
distribution function contained in $G^K$ now may not be thermal.
However, the more important difference is that the gap equation
(\ref{Gap}) is now complex and gives two equations for two
unknowns: the order parameter $\psi$ and the frequency $\mu_S$.

The common oscillation frequency $\mu_S$ would in thermal equilibrium
be the system's chemical potential, considered as a control parameter,
adjusted to match the required density, and the (real) gap equation
determines only $\psi$.
Here, because different baths have different chemical potentials, the
system is not in chemical equilibrium with either bath, so both
$\mu_S$ and $\psi$ must be found from the gap equation.
The density, which can be found given $\psi$ and $\mu_S$,
is set by the relative strength of the pump and decay.

Thus the real part of the gap equation is analogous to the gap equation for
closed equilibrium system, where the right hand side describes polarisation
due to nonlinear susceptibility.
By considering the existence of pumping and decay, one also introduces the
imaginary part, which describes how the gain balances the decay (as in lasers)
but now in the strongly coupled exciton-photon system.
If one were to instead consider the equilibrium theory, and merely add decay
rates, one could not a priori guarantee that the fluctuation spectrum would be
gapless, as should arise from spontaneous symmetry breaking.
 By ensuring that gain and decay balance,
the fluctuation spectrum above the ground state which
satisfies both the real and imaginary parts of the gap equation, will indeed be
gapless.
By connecting the equilibrium self consistency condition (gap equation, Gross
Pitaevskii equation), and the laser rate equation, Eq.~(\ref{Gap}) puts the
condensate and the laser\cite{Haken:RMP} in the same framework
and so allows study of the crossover and the relation between the two.

Using (\ref{GMF5}) the mean-field equation becomes
\begin{multline}
(\omega_0-\mu_S-i\kappa)\psi_{f} = \\ \sum_{\alpha} \int \frac{d \nu}{2 \pi}
\psi_{f} g^2 \gamma
\frac{(F_a+F_b)\nu+(F_b-F_a)(\tilde{\epsilon}_\alpha+i\gamma)}
{[(\nu-E_{\alpha})^2+\gamma^2][(\nu+E_{\alpha})^2+\gamma^2]}.
\label{gap2}
\end{multline}
Note that $\gamma$ appears both in the denominator (as it gives rise to the
dephasing) and in the numerator (it gives rise to pumping).

As in thermal equilibrium, the normal state $\psi_f=0$ is always a solution of
Eq.~(\ref{gap2}), but for some range of parameters there is also a condensed
$\psi_f \ne 0$ solution. For $\psi_f \ne 0$ the final form of the gap
equation is 
\begin{multline}
\tilde{\omega}_0-i\kappa=  \\ g^2 \gamma \sum_{\alpha} \int \frac{d
  \nu}{2 \pi}  
\frac{(F_a+F_b)\nu+(F_b-F_a)(\tilde{\epsilon}_\alpha+i\gamma)}
{[(\nu-E_{\alpha})^2+\gamma^2][(\nu+E_{\alpha})^2+\gamma^2]}.
\label{sp2}
\end{multline}
Now for a given set of parameters we can solve the real and imaginary parts of
this equation to determine the coherent photon field $\psi_f$
and its oscillation frequency $\mu_S$.

We can reduce the number of parameters in our theory by measuring
energies in units of the exciton-photon coupling $g$ and, noting that
our equations have made no assumption about the origin of energies,
taking $\epsilon_0$ as some reference energy, such as the bottom
of the exciton band.
The independent parameters in our theory are then the distribution of
exciton energies [i.e. $\sum_{\alpha} \to \int d\epsilon
\nu_S(\epsilon)$], the detuning of the photon from the reference point
$\Delta=\omega_0-2\epsilon_0$, the pumping bath chemical potential
$\mu_B - \epsilon_0$, the pumping (decoherence) strength $\gamma$, and
the coupling to the decay bath $\kappa$.

Having found the self consistent oscillation frequency $\mu_S$
and coherent field, one can then calculate the excitonic density and
polarisation.
The polarisation, {\it i.e.} $\langle a^{\dagger} b \rangle$ (where
$|\langle a^{\dagger} b \rangle|^2$ also gives the number of condensed
fermion pairs - condensed excitons) follows directly from the gap
equation, so the magnitude of polarisation is given by
$\sqrt{\tilde{\omega}^2 + \kappa^2} \psi_f$.
The excitonic density is given by:
\begin{multline}
\sum_{\alpha}
\frac{1}{2}\left(b^{\dagger}_{\alpha}b_{\alpha}-a^{\dagger}_{\alpha}a_{\alpha}\right)=
\\ \frac{1}{4}i \sum_{\alpha} \int\frac{d\nu}{2\pi}\left( G^K_{aa}(\nu) -
G^K_{bb}(\nu) \right) =
\frac{\gamma}{2} 
\sum_{\alpha} \int \frac{d \nu}{2\pi}  \\
\hspace{-0.03\textwidth}\times  \frac{
      (F_b-F_a)\left[
        g^2|\psi|^2 - \nu^2 -\tilde{\epsilon}_\alpha^2 - \gamma^2  \right] - 
      (F_b+F_a)2\nu\tilde{\epsilon}_\alpha}
    {[(\nu-E_{\alpha})^2+\gamma^2][(\nu+E_{\alpha})^2+\gamma^2]}. 
\label{spd1}
\end{multline}
Since our choice of bath populations in
Eq.~(\ref{FaFb}) implies that the empty state corresponds to
$\langle b^{\dagger} b \rangle=0$, $\langle a^{\dagger} a \rangle=1$, 
it will be convenient to shift the exciton density so that
the empty state corresponds to zero density, thus:
\begin{equation}
  \label{eq:rhofermi}
  \rho_{\mathrm{exciton}} = \sum_{\alpha}
  \frac{1}{2}\left(
    1+ b^{\dagger}_{\alpha}b_{\alpha}-a^{\dagger}_{\alpha}a_{\alpha}
  \right).
\end{equation}
In the limit that the temperature of the pumping bath goes to zero, one
can perform the various integrals in Eq.~(\ref{sp2}) and
Eq.~(\ref{spd1}) in terms of elementary functions.  
These forms are presented in Appendix~\ref{sec:gap-equation-at}.

\subsection{$\gamma = 0, \kappa=0$ limit}
\label{sec:gamma-=-0}

In order to understand the meaning of the gap equation, and the
connection to condensation in a closed equilibrium system, it is
instructive to take the limit $\gamma \to 0, \kappa \to 0$ in
Eq.~(\ref{sp2}).
This will also provide a consistency check of the non-equilibrium
theory as it should recover the equilibrium limit as the coupling to
the environment  approaches zero. 
The real part of Eq.~(\ref{sp2}) can be rewritten as:
\begin{multline*}
\hspace{-0.02\textwidth}\omega_0 - \mu_S = \frac{g^2}{4E} \int \frac{d
\nu}{2 \pi} \left[\frac{\gamma}{(\nu-E)^2+\gamma^2} -
\frac{\gamma}{(\nu+E)^2+\gamma^2}\right]
\\
\times
\left[(F_a(\nu)+F_b(\nu)) + (F_b(\nu)-F_a(\nu))
\frac{\tilde{\epsilon}_\alpha}{\nu}\right].
\end{multline*}
From the definition of the $\delta$ function we have 
\begin{displaymath}
\lim_{\gamma \to 0} \frac{\gamma}{(\nu-E)^2+\gamma^2} = \pi
\delta(\nu-E) 
\end{displaymath}
and so, using $F_a(-\nu) = - F_b(\nu)$,
the real part of the gap equation reduces to
\begin{multline}
  \omega_0-\mu_S= 
  \frac{g^2}{4E}
  \left[
    F_b(E) +
    F_a(E)
  \right] \\
  + \frac{g^2\tilde{\epsilon}}{4E^2}
  \left[
    F_b(E) -
    F_a(E)
  \right].
\label{T=0re}
\end{multline}
Similarly, the imaginary part of the gap equation can be rearranged as:
\begin{equation}
  \frac{\kappa}{\gamma} = 
  \frac{g^2}{4E^2}
  \left[
    F_a(E) -
    F_b(E)
  \right].
\label{T=0im}
\end{equation}
Let us consider the limit where $\kappa/\gamma \to 0$, i.e.  coupling
to the photon bath vanishes faster, and so the distribution will be
set by the pumping bath.
Then the left hand side of Eq.~(\ref{T=0im}) is zero and so one requires
$F_a(E)=F_b(E)$.
Using the gauge transformed versions of the thermal distribution
functions in Eq.~(\ref{FaFb}), this condition becomes $\tilde{\mu}_B=0$,
i.e. that $\mu_S=2\mu_B$.
Putting this solution into (\ref{T=0re}) we recover the
equilibrium gap equation at a temperature $T$ set by the pumping bath:
\begin{displaymath}
\tilde{\omega}_0= \frac{g^2}{2E}\rm{tanh}\frac{\beta}{2}E.
\end{displaymath}
This limit provides a reassuring test of the formalism, and also
supports the interpretation that the real part
of the gap equation connects the order parameter with
non-linear susceptibility, while the imaginary part describes the balance of
gain and decay, and so controls $\mu_S$ and the particle density in the
system.

\section{Second Order Fluctuations and Stability of Solutions}
\label{sec:second-order-fluct}

Having found the self-consistency condition, considering the
possibility of uniform condensed solutions, we next consider the
stability of such solutions.
The consideration of stability is important firstly since, as
discussed above, $\psi_f=0$ is always a solution of Eq.~(\ref{Gap}),
so one must determine which of the normal and condensed solutions is
stable, and secondly because we considered only spatially
homogeneous fields, with a single oscillation frequency, so one may
find that neither $\psi_f=0$ nor our ansatz of Eq.~(\ref{ansatz}) is
stable, suggesting more interesting behaviour.
There is an important difference in interpretation of the saddle point
equation between the closed-time-path path-integral formalism
used here, and the imaginary-time path-integral in thermal
equilibrium.
In the imaginary time formalism, extremising the action corresponds to
finding configurations which extremise the free energy; thus,
stable solutions correspond to a minimum of free energy, and
unstable to local maxima.
Here in contrast, for a classical saddle point (i.e.
  $\psi_q=0$), the action is always $S=0$, and the saddle point
condition corresponds to configurations for which nearby paths add in
phase.
Thus, in order to study stability one must directly investigate
fluctuations about our ansatz, and determine whether such
fluctuations grow or decay.

In considering the question of stability, we will first discuss stability of
the normal state, which is instructive as it shows how the question of whether
fluctuations about the non-equilibrium steady-state grow or
decay is directly related to the instability expected in thermal
equilibrium systems when the chemical potential goes above a bosonic mode.
We will then turn to the spectrum of fluctuations about our
condensed ansatz.
While we will discuss here whether such fluctuations are stable or
unstable, we will defer until Sec.~\ref{sec:fluct-cond-state} the
evaluation of correlation functions associated with these
fluctuations.
This is because, as discussed there, these fluctuations include phase modes,
and phase fluctuations may become large.
It is therefore insufficient to only expand to second order in
fluctuation fields, but one must instead reparameterise $\psi = \sqrt{\rho_0 +
\pi}e^{i\phi}$, and then describe the correlation functions of $\psi$ in terms
of those of phase $\phi$ and amplitude $\pi$,
including the effects of $\phi$ to all orders.
Such a complication is however not needed in order to study whether
fluctuations are stable or not, and so it is reasonable to postpone such a
treatment, and consider an expansion in terms of $\psi = \psi_0 + \delta \psi$
 to second order in $\delta \psi$ instead.

Thus, to find the spectrum of fluctuations, we consider the effective
action governing fluctuations about either $\psi=\psi_0$ or about
$\psi=0$.
Considering the effective action in Eq.~(\ref{totalS}), and expanding
to second order in $\delta \psi$, one finds a contribution
from the effective photon action, and a contribution from
expanding the trace over excitons.
This latter contribution can be found by writing
\begin{math}
  G_{\alpha,\alpha}^{-1}
  =
  (G^{\mathrm{sp}}_{\alpha,\alpha})^{-1}
  +
  \delta G_{\alpha}^{-1},
\end{math}
where $G^{\mathrm{sp}}$ is the saddle point fermionic Green's function,
which depends on the value of the saddle point field $\psi_f$, as
given in Eq.~(\ref{GMF1}), (\ref{GMF3}) and~(\ref{GMF5}), and
the contribution of fluctuations $\delta{G}_{\alpha}^{-1}$ is given by:
\begin{multline*}
\delta G^{-1}= \frac{-1}{\sqrt{2}}(g\delta \bar{\psi_q}
\sigma_{-} + g\delta \psi_q \sigma_{+}
)\sigma_{0}^{\textsc{k}} + \\  \frac{-1}{\sqrt{2}}(g\delta \bar{\psi}_{cl}
\sigma_{-} + g\delta \psi_{cl} \sigma_{+}
)\sigma_{1}^{\textsc{k}}.      
\end{multline*}
Thus, one can expand the action as:
\begin{multline*}
  -i\sum_{\alpha}{ \rm Tr ln}
  \left[ G_{\alpha,\alpha}^{-1} \right]
  =
  (-i)
  \sum_{\alpha}{ \rm Tr ln}
  \left[ (G^{\mathrm{sp}}_{\alpha,\alpha})^{-1} \right]
  \\+
  (-i)
  \sum_{\alpha}{ \rm Tr}
  \left [G^{\mathrm{sp}}_{\alpha,\alpha} \delta G_{\alpha}^{-1} \right]
  \\+
  (-i)(-\frac{1}{2})
  \sum_{\alpha}{ \rm Tr}
  \left[
    G^{\mathrm{sp}}_{\alpha,\alpha} \delta G_{\alpha}^{-1}
    G^{\mathrm{sp}}_{\alpha,\alpha} \delta G_{\alpha}^{-1}
  \right].
\end{multline*}
In this expansion, we have retained only the terms diagonal in
site index; i.e. neglected any bath induced interaction between
different exciton sites.  
Such bath induced interactions should be small for small $\gamma$, and
their inclusion would considerably complicate the formalism.
Such an approach is also equivalent to considering a separate set of
baths for each disorder localised state $\alpha$.

Because, in the presence of a coherent field, the effective
action can contain terms like $\delta \psi \delta \psi$ and
$\delta \bar{\psi} \delta \bar{\psi}$, it is convenient
to introduce a Nambu structure of photon fields.
Thus, the photon fluctuations are described by a $2\times2=4$
component vector, with one factor of $2$ from the Keldysh
structure, and one from the Nambu structure, hence:
\begin{equation}
  \label{eq:nambu-fluct}
  \delta \Lambda
  =\left(
    \begin{array}{c}
      \delta \psi_{cl}(\omega) \\
      \delta \bar{\psi}_{cl}(-\omega) \\
      \delta \psi_q(\omega) \\
      \delta \bar{\psi}_q(-\omega) 
    \end{array} \right),
\end{equation}
in terms of which the action for fluctuations $\delta S_f$ is:
\begin{displaymath}
  \delta S_f =  \int \frac{d \omega}{2 \pi} 
  \delta \bar{\Lambda}(\omega)
  \left(
    \begin{array}{cc}
      0 & \left[{\cal D}^{-1}\right]^{A}
      \\
      \left[{\cal D}^{-1}\right]^{R} & \left[{\cal D}^{-1}\right]^{K}
    \end{array} \right) 
  \delta \Lambda(\omega).
\end{displaymath}
For convenience later, we shall introduce the notation:
\begin{equation}
  \label{eq:define-K}
  \left[{\cal D}^{-1}\right]^{R/A/K}
  =
  \left(
    \begin{array}{cc}
      K^{R/A/K}_1& K^{R/A/K}_2
      \\
      \\
      K^{R/A/K}_3& K^{R/A/K}_4
    \end{array} \right).
\end{equation}
By definition we have that: $\left[{\cal
    D}^{-1}\right]^{A}=\left(\left[{\cal
      D}^{-1}\right]^{R}\right)^{\dagger}$, and in addition the Nambu
structure implies certain symmetries between the elements of $[{\cal
  D}^{-1}]^{R/A/K}$, which together can be written as:
\begin{equation}
  \label{eq:dynamic-symmetries}
  \begin{array}{l@{\mbox{}=\mbox{}}l@{\mbox{}=\mbox{}}l@{\mbox{}=\mbox{}}l}
  K^{R}_{1}( \omega)        &  K^{A}_{1}( \omega)^{\ast} &
  K^{R}_{4}(-\omega)^{\ast} &  K^{A}_{4}(-\omega) 
  \\[1.2ex]
  K^{R}_{2}( \omega)        &  K^{A}_{2}(-\omega) &
  K^{R}_{3}(-\omega)^{\ast} &  K^{A}_{3}( \omega)^{\ast}
  \end{array}
\end{equation}
\begin{equation}
  \label{eq:keldysh-symmetries}
  \begin{split}
    K^{K}_{2}(\omega)         &= -K^{K}_{3}( \omega)^{\ast} 
    =  K^{K}_{2}(-\omega) 
    \\
    K^{K}_{1}(\omega)         &=  K^{K}_{4}(-\omega) 
  \end{split}
\end{equation}

Introducing the compact notation:
\begin{displaymath}
  [f \ast g]_{\omega} =
  \int \frac{d\nu}{2\pi} f(\nu) g(\nu-\omega)
\end{displaymath}
we may thus write:
\begin{widetext}
  \begin{multline}
    \left[{\cal D}^{-1}\right]^{R}(\omega,\vect{p})
    =
    \frac{1}{2} 
    \left(
      \begin{array}{cc}
        \omega -\tilde{\omega}_{\vect{p}}+i\kappa  & 0
        \\
        \\
        0 & -\omega -\tilde{\omega}_{\vect{p}}-i\kappa
      \end{array} \right) 
    +
    i\frac{g^2}{4} 
    \left(
      \begin{array}{ccc}
        G^R_{bb} \ast G^K_{aa}+
        G^K_{bb} \ast G^A_{aa} & \ \
        G^R_{ba} \ast G^K_{ba}+
        G^K_{ba} \ast G^A_{ba}
        \\
        \\
        G^R_{ab} \ast G^K_{ab}+
        G^K_{ab} \ast G^A_{ab} & \ \ 
        G^R_{aa} \ast G^K_{bb}+
        G^K_{aa} \ast G^A_{bb}
      \end{array} \right)_\omega,
    \label{eq:retarded-inverse-gf}
  \end{multline}
  and
  \begin{multline}
    \left[{\cal D}^{-1}\right]^{K}(\omega,\vect{p})=
    \frac{1}{2} 
    \left(
      \begin{array}{cc}
        2i\kappa F_{\psi}(\omega+\mu_S)  & 0
        \\
        \\
        0 & 2i\kappa F_{\psi}(-\omega+\mu_S) 
      \end{array} \right)
    +\\
    i\frac{g^2}{4} 
    \left(
      \begin{array}{ccc}
        G^K_{bb} \ast G^K_{aa}+
        G^R_{bb} \ast G^A_{aa}+
        G^A_{bb} \ast G^R_{aa} & \  \
        G^K_{ba} \ast G^K_{ba} +
        G^R_{ba} \ast G^A_{ba} +
        G^A_{ba} \ast G^R_{ba}
        \\
        \\
        G^K_{ab} \ast G^K_{ab} +
        G^R_{ab} \ast G^A_{ab} +
        G^A_{ab} \ast G^R_{ab}  &  \  \
        G^K_{aa} \ast G^K_{bb} +
        G^R_{aa} \ast G^A_{bb} +
        G^A_{aa} \ast G^R_{bb}
      \end{array} \right)_\omega
    \label{eq:keldysh-inverse-gf}
  \end{multline}
\end{widetext}

\subsection{Normal state excitation spectra and distributions}
\label{sec:norm-state-excit}

The excitation spectrum can be found from the poles of the fluctuation Green's
function, i.e. from the zeros of  $\det\left[{\cal D}^{-1}\right]^R$.
To extract the occupation of the spectrum, one can extract the
boson distribution function via
\begin{displaymath}
  {\cal D}^K 
  =
  - {\cal D}^R \left[{\cal D}^{-1}\right]^{K} {\cal D}^A
  =
  {\cal D}^R F_S - F_S {\cal D}^A,
\end{displaymath}
where simply
${\cal D}^{R/A} = \left[ \left[{\cal D}^{-1}\right]^{R/A} \right]^{-1}$.
Whilst in general these are $2\times 2$ matrices in Nambu space, in the normal
state this structure is redundant, and so the distribution function is the
diagonal constant matrix $F_S = 2 n_S + 1$, where $n_S$ describes the
occupation of the modes.
Alternatively, one can invert the Keldysh rotation in
order to find the physical Green's functions,
\begin{equation}
  \label{eq:inv-keldysh-1}
  {\cal D}^{<,>} = \frac{1}{2}
  \left(
    {\cal D}^{K} 
    \mp \left[
      {\cal D}^{R} - {\cal D}^{A}
    \right]
  \right),
\end{equation}
which as discussed in Sec.~\ref{sec:path-integr-form} relate directly
to the luminescence, ${\cal L}(\omega,\vect{p}) = i {\cal
  D}^{<}(\omega,\vect{p})/2\pi$, and absorption ${\cal
  A}(\omega,\vect{p}) = i {\cal D}^{>}(\omega,\vect{p})/2\pi$.
Still, assuming the normal state, so that the Nambu structure is
redundant, these become:
\begin{align*}
  {\cal L}(\omega,\vect{p}) &= n_{S}(\omega) 
  \Im\left( -\frac{{\cal D}^R(\omega,\vect{p})}{\pi} \right),
  \\
  {\cal A}(\omega,\vect{p}) &= \left(n_{S}(\omega)  + 1 \right)
  \Im\left( -\frac{{\cal D}^R(\omega,\vect{p})}{\pi} \right).
\end{align*}
While this form illustrates how the spectral weight and occupation can
be separately extracted from the luminescence and absorption, in order
to study these quantities it is more helpful to write them in terms of
the components, $K_1^{R,K}$ of the inverse Green's function.
In the normal state, there are no anomalous (off diagonal in Nambu space)
contributions, and so $K_2^{R/A/K} = K_3^{R/A/K} = 0$.
Thus, the normal state luminescence, absorption, and distribution
functions are given by:
\begin{align}
  \label{eq:lum-abs-K}
  \left({\cal L},{\cal A}\right)(\omega,\vect{p})
  &=
  \frac{%
    -i K_1^K(\omega) \mp 2 \Im \left[ K_1^R(\omega) \right]
  }{4\pi   |K_1^R(\omega,\vect{p})|^2},
  \\
  \label{eq:distribution-K}
  F_S(\omega) &=
  \frac{-i K_1^K(\omega)}{2 \Im \left[ K_1^R(\omega) \right]}. 
\end{align}

\begin{figure}[htpb]
  \centering
  \includegraphics[width=0.98\columnwidth]{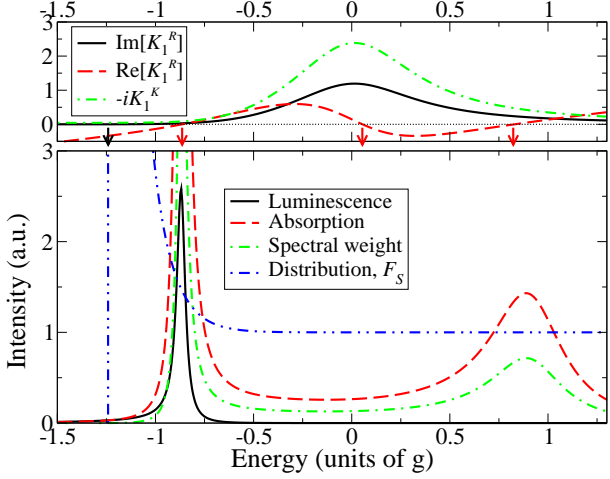}
  \caption{(Color online) Relation between real and imaginary parts of
    $K_1^R$ and $K_1^K$ and the luminescence, absorption, spectral
    weight and distribution function.  Plotted for $\gamma=0.2g,
    \kappa=0.02 g, T=0.1g, \mu_B=-0.5g$.  (cf
    Fig.~\ref{fig:trace-ns-zeros}) The upper panel shows the real and
    imaginary parts of $K_1^R$, and their zeros marked by arrows.  The
    lower panel shows how these lead to peaks in the luminescence
    spectrum, and how the zero of $\Im[K^1_R]$ defines divergence of
    the distribution function.}
  \label{fig:normal-luminescence}
\end{figure}

Let us now discuss what can be understood in general from the form of
these equations, and then illustrate this discussion with the simple
case $\gamma \ll T$.
From the difference of luminescence and absorption in
Eq.~(\ref{eq:lum-abs-K}), one can identify a spectral weight:
\begin{equation}
  \label{eq:spectral-K}
  2 \pi S(\omega,\vect{p})
  =
  \frac{%
    \Im \left[ K_1^R(\omega) \right]
  }{\Re\left[ K_1^R(\omega,\vect{p}) \right] ^2 + \Im\left[ K_1^R(\omega) \right] ^2}
\end{equation}
thus, if the imaginary part of $K_1^R(\omega)$ is a smooth
function of omega, then one will have almost Lorentzian peaks of the
spectral weight at values $\omega^{\ast}$ where $\Re\left[
  K_1^R(\omega^{\ast}) \right]=0$.
The width of these peaks, i.e. the linewidth, is then given by
$\Im\left[ K_1^R(\omega^{\ast}) \right]$.
Thus, the imaginary part plays one role as determining the linewidth.
It also plays a second role, since from Eq.~(\ref{eq:distribution-K}),
a zero of the imaginary part causes the distribution to diverge;
however, at these same points Eq.~(\ref{eq:spectral-K}) implies
the spectral weight vanishes, so the number of photons does not
diverge.
Since a Bose distribution would diverge at the chemical potential, we
can use this as a definition of an effective boson chemical potential,
so $\Im\left[ K_1^R(\mu^{\mathrm{eff}}) \right]=0$.
These results are illustrated in Fig.~\ref{fig:normal-luminescence},
which show the luminescence, absorption, spectral weight, and
distribution function against the real and imaginary parts of $K_1^R$
and $K_1^K$.

From the above, it is clear that the form of $\Im\left[
  K_1^R(\omega) \right]$ as well as $K_1^K(\omega)$
conspire to set the effective photon distribution.
Using the expressions in Eq.~(\ref{GMF1}), (\ref{GMF3}), and~(\ref{GMF5}),
and for the moment restricting to the case $\epsilon_{\alpha}=\epsilon$
we may write:
\begin{widetext}
\begin{align}
  \label{eq:K1K}
  -i K_1^K(\omega)
  &= \kappa F_{\Psi}(\omega)
  + \frac{g^2}{4}
  \left[
    2 \Re
    \int \frac{d \nu}{2\pi} 
    \frac {1}{
      (\nu - \epstil + i \gamma)(\nu - \omega + \epstil - i\gamma)
    }
    -
    4\gamma^2 \int \frac{d \nu}{2\pi} 
    \frac { F_b(\nu) F_a(\nu - \omega)}{
      \left[(\nu-\epstil)^2 + \gamma^2\right]
      \left[(\nu-\omega+\epstil)^2 + \gamma^2\right]
    }
  \right]
  \\
  2 \Im\left[K_1^R(\omega)\right]
  &=
  \kappa
  + g^2 \gamma^2
  \int \frac{d \nu}{2\pi} 
  \frac { F_b(\nu) - F_a(\nu - \omega)}{
    \left[(\nu-\epstil)^2 + \gamma^2\right]
    \left[(\nu-\omega+\epstil)^2 + \gamma^2\right]
  }.
\end{align}
\end{widetext}
For the case of pumping baths being individually in thermal
equilibrium, one may get some insight into how the pump and
decay baths compete to set the systems distribution.
In the limit $\gamma \ll T$, where $T$ is the temperature of the
pumping bath, the distribution functions $F_{b/a}(\nu)$ are smooth,
while the denominators lead to sharp peaks, of width $\gamma$.
One can then approximate the integrals by assuming that over each
Lorentzian peak, the distribution function takes its value at
the maximum of that peak, and so:
\begin{equation}
  \label{eq:distribution}
  F_S(\omega) = 
  \frac{\displaystyle
    \kappa F_{\Psi}(\omega)
    +
    \frac{
      g^2 \gamma \left( 1- F_b(\epstil) F_a(\epstil - \omega) \right) 
    }{
      (\omega-2\epstil)^2 + 4 \gamma^2
    }
  }{\displaystyle
    \kappa
    +
    \frac{
      g^2 \gamma \left( F_b(\epstil) - F_a(\epstil - \omega) \right) 
    }{
      (\omega-2\epstil)^2 + 4 \gamma^2
    }
  }
\end{equation}

From this one can see immediately two trivial limits. 
If $\gamma=0$ or if $g=0$, then there is no influence of the
pumping bath and so $F_S(\omega)=F_{\Psi}(\omega)$, i.e the
photon distribution in the system is the same as the distribution of
bulk modes outside the cavity.
Similarly, if $\kappa=0$, the photon bath has no effect, and
\begin{displaymath}
  F_S(\omega) = 
  \frac{%
     1- F_b(\epstil) F_a(\epstil - \omega) 
    }{%
     F_b(\epstil) - F_a(\epstil - \omega)  
    }.
\end{displaymath}
Thus, as one might expect, if the fermions are in thermal
equilibrium with $F_{b,a}(\nu) = F(\nu \mp \tilde{\mu}_B)$
where $F(\nu)=\tanh(\beta \nu/2)$, then by using a standard
hyperbolic trigonometric identity, this gives a thermal Bose
distribution for the photons, with the same temperature, but twice the
chemical potential, as expected since one boson corresponds to two
fermions:
\begin{align*}
  F_S(\omega) 
  &= 
  \coth \left[
    \frac{\beta}{2} \left(\epstil - \tilde{\mu}_B\right) - 
    \frac{\beta}{2} \left(\epstil - \omega + \tilde{\mu}_B \right)
  \right]
  \\
  &=  \coth \left[ \frac{\beta}{2} \left(\omega - 2
  \tilde{\mu}_B\right) \right]. 
\end{align*}
The above expressions have been written after the gauge
transformation described following Eq.~(\ref{ansatz}).
Of course, in the normal state, such a gauge transform has no effect,
since it just corresponds to an arbitrary shift of the origin for
measuring energies, but we use the transformed notation for consistency
with the condensed case.

More generally, the two distributions compete to control the photon
distribution, which in general will not be thermal even if the
baths are individually thermal, because they have different chemical
potentials and temperatures.
It is clear from Eq.~(\ref{eq:distribution}) that the effect of the
pumping bath is largest near $\omega=2\epstil$, and far
from this value, both numerator and denominator are instead dominated
by the photon bath.
Physically, this means that the effect of the pumping bath is
only important at energies where the photons are nearly resonant with,
and so couple strongly to, the excitons.

\subsection{Instability of the normal state above the transition}
\label{sec:inst-norm-state}

\begin{figure}[htpb]
  \centering
  \includegraphics[width=0.98\columnwidth]{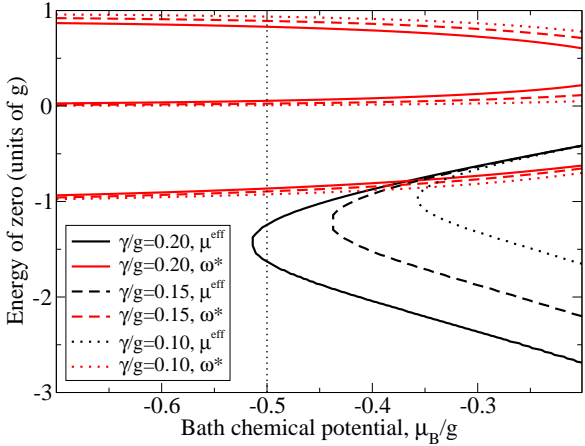}
  \caption{(Color online) Energy of zeros of the real and imaginary
    part of $K_1^R$ as a function of $\mu_B$, demonstrating how
    $\mu^{\mathrm{eff}}$, the zero of $\Im(K_1^R)$ approaches
    $\omega^{\ast}$, a zero of $\Re[K_1^R]$ at the transition.
    Results are plotted for $\kappa=0.02g, T_B=0.1g$, and three values
    of $\gamma$ as indicated.  The dotted vertical line marks the
    locations of the trace plotted in
    Fig.~\ref{fig:normal-luminescence}}
  \label{fig:trace-ns-zeros}
\end{figure}

The discussion in the previous section, which defined $\mu^{\mathrm{eff}}$ by
zeros of the imaginary part of $K_1^R$, and $\omega^{\ast}$ by zeros
of the real part allows one to understand the instability
of the normal state.
It can be seen that the gap equation, (\ref{sp2}), if evaluated at $\psi_f=0$,
is equivalent to the condition $K_1^R(\omega=0,\vect{p}=0)=0$,
(measuring $\omega$ 
relative to $\mu_S$).
This can be understood physically by seeing that the vanishing of
$K_1^R(\omega=0,\vect{p}=0)$ implies there is a zero mode, corresponding to
global phase rotations, as one expects in a broken symmetry system.
Thus, this condition implies that there is a frequency at which both
real and imaginary parts simultaneously vanish; i.e.  the gap equation
is the condition that $\mu^{\mathrm{eff}}=\omega^{\ast}$, the
effective chemical potential reaches the bottom of the normal mode
spectrum.
One can say ``bottom of the spectrum'' since the ${\vect{p}}$
dependence only enters the real part of $K_1^R$, and $\omega^{\ast}$
will increase as ${\vect{p}}$ increases, thus if $\omega^{\ast}_{\vect{p}=0} <
\mu^{\mathrm{eff}}$, then there will be a non-zero ${\vect{p}}$ for which
$\omega^{\ast}_{\vect{p}} = \mu^{\mathrm{eff}}$.
Thus, the existence of a non-trivial solution to the gap equation can
still be understood as a ``chemical potential'' reaching the bottom of
the band\cite{Zimmermann06}, even in this non-equilibrium context, as
is illustrated in Fig.~\ref{fig:trace-ns-zeros}.

It is also possible to connect the effective chemical
potential reaching the bottom of the band to instability of the normal state,
i.e.  fluctuations growing in time.
Let us consider poles, $\xi_{\vect{p}}$ of the retarded Green's function, i.e.
zeros of $K_1^R(\xi_{\vect{p}},\vect{p})$.
If these poles have negative imaginary parts they correspond to
fluctuations that decay in time, and if positive, to growing
fluctuations;  thus stability requires the imaginary part
to be always negative.
It is clear that at large enough momenta, the Green's function is
that of bare photons, and is stable.
Thus, if there are to be unstable modes, then there must be some
${\vect{p}}$ value at which the imaginary part of the poles goes from negative
to positive.
For reasonable systems, where the linewidth is a smooth function of
momentum, this means the imaginary part must go through zero.
A zero of $\Im[\xi_{\vect{p}}]$ means there is a real frequency which
satisfies $K_1^R(\xi_{\vect{p}},\vect{p})=0$.
However, the existence of a real frequency satisfying this condition
was, as discussed previously, exactly the gap equation at $\psi=0$.
Thus, if $\xi_{\vect{p}}=\omega^{\ast}_{\vect{p}} =
\mu^{\mathrm{eff}}$ for some $|\vect{p}|=p_c$, then for $|\vect{p}| < p_c$ one
will find positive imaginary parts.
To illustrate this, consider a linear expansion in $\omega$, so
that:
\begin{displaymath}
  K_1^R(\omega,\vect{p}) \simeq 
  (\omega - \omega^{\ast}_{\vect{p}}) + 
  i \alpha (\omega - \mu^{\mathrm{eff}})
  \simeq C (\omega - \xi_{\vect{p}}),
\end{displaymath}
then one finds, $\Im \xi_{\vect{p}} \propto (\mu^{\mathrm{eff}} -
\omega^{\ast}_{\vect{p}})$.

\begin{figure}[htpb]
  \centering
  \includegraphics[width=0.98\columnwidth]{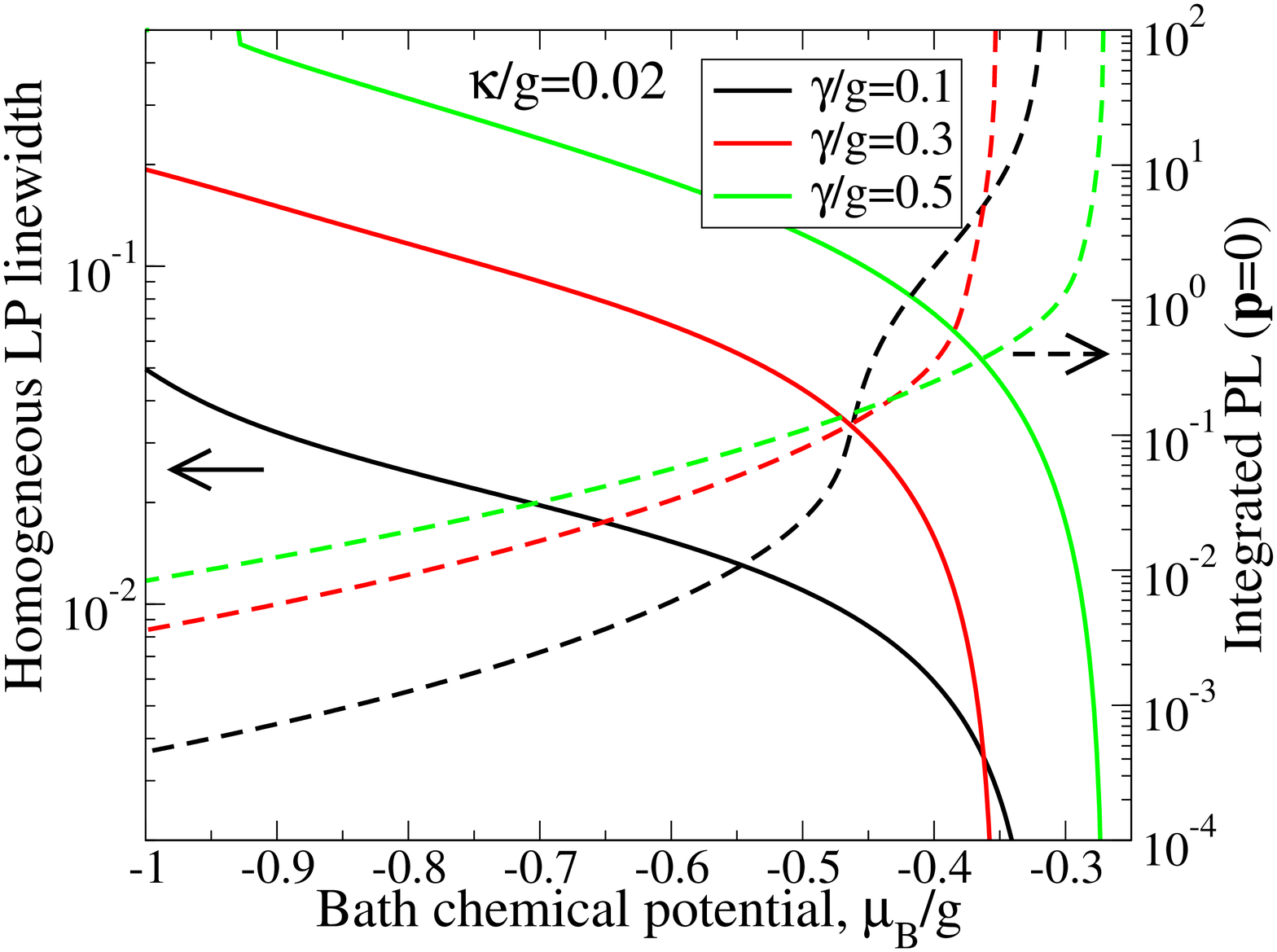}
  \caption{(Color online) Linewidth of the lower mode (solid,
    left axis), and energy integrated luminescence
    $\vect{p}_{\parallel}=0$ (dashed, right axis) in the normal state
    as a function of pumping bath chemical potential, as the phase
    boundary is approached.  Results are shown for two different
    dephasing parameters $\gamma$ with $\kappa=0.02g$ for all three.}
  \label{fig:linewidth}
\end{figure}

Two more important connections can be drawn from the relation between
poles of the retarded Green's function, the distribution, and the
gap equation.
The first is that, 
as for any second-order phase transition,
approaching the phase transition from the normal side, the
fluctuation Green's function describes a susceptibility which diverges
at the transition.
The second relates to the dual role that $\Im \left[
  K_1^R(\omega^{\ast}_{\vect{p}})\right]$ played as the linewidth.
As one approaches the phase boundary, at which real and imaginary
parts both have zeros, one must have that the effective linewidth
vanishes.  
These points are illustrated in Fig.~\ref{fig:linewidth}.
Note however that $\Im \left[ K_1^R(\omega)\right]$ is of course not a
constant, and so there will be some non-trivial lineshape, but a
linewidth defined by full width half maximum will vanish on
approaching the condensed state, as a peak develops at $\omega=0$.
The vanishing of homogeneous linewidth at the transition is a manifestation of
diverging susceptibility in an {\it infinite} system.
{\it Finite} system size is expected to smear out this divergence and
result in the homogeneous linewidth remaining non-zero, but still
having a minimum near the transition.
Additionally inhomogeneous broadening of exciton energies will add to
the linewidth measured in experiments.

\subsection{Fluctuations in condensed state - stability and collective modes}
\label{sec:fluct-cond-state-1}

From the previous section we conclude that when there is a non-trivial
solution to the gap equation, the normal state is unstable.
We wish now to determine whether our ansatz of Eq.~(\ref{ansatz}) is
stable.
As discussed above, if there were a region with unstable modes (i.e.
positive imaginary parts of poles), then this would lead to the
existence of a true pole at real omega, at the boundary of the
unstable region.
Making use of the symmetries in Eq.~(\ref{eq:dynamic-symmetries}), for
the condensed case, poles of the retarded Green's function correspond
to solutions of
\begin{equation}
  \label{eq:denom-zeros}
  K_1^R(\omega,\vect{p}) K_1^R(-\omega,\vect{p})^{\ast}
  -
  K_2^R(\omega) K_2^R(-\omega)^{\ast} = 0.
\end{equation}

Unfortunately this expression is not simple, and numerical
evaluation would be necessary to trace the behaviour of all
zeros as a function of momentum.
However, in order to understand the stability, we can instead consider
separately zeros of the real and imaginary parts of
Eq.~(\ref{eq:denom-zeros}).
If zeros of these two parts coincide for some $p_c$ , there is a real
pole, and thus instability for $|\vect{p}|<p_c$.
It is clear the imaginary part should have a zero at $\omega=0$
(measuring frequency from the common oscillation frequency $\mu_S$),
as the imaginary part of Eq.~(\ref{eq:denom-zeros}) is an odd function
of $\omega$.
This zero physically corresponds to the divergence of the distribution
function at $\omega=0$.  
Numerical investigation suggests that this is the only zero of the
imaginary part.
Thus, we are interested in zeros of the real part, evaluated at
$\omega=0$, but arbitrary ${\vect{p}}$.

It is clear there is a zero at $\omega=0, \vect{p}=0$, corresponding to the
symmetry under global phase rotations, but being at $\vect{p}=0$ this does
not lead to instability.
From this pole, or alternatively working directly from the definitions
of $K^R$ in Eq.~(\ref{eq:retarded-inverse-gf}),
 and the gap equation~(\ref{sp2}),
one can show that $K_1^R(\omega=0,\vect{p}=0)=K_2^R(\omega=0)$.
Thus, writing $A=\Re\left[K_1^R(\omega=0,\vect{p}=0)\right]$,
instability occurs 
if there is a non-zero ${\vect{p}}$ solution of:
\begin{displaymath}
  \left( A - \frac{1}{2} \frac{\vect{p}^2}{2m_{\mathrm{ph}}}\right)^2 - A^2
  =
  \frac{1}{4}\frac{\vect{p}^2}{2m_{\mathrm{ph}}} \left(
    \frac{\vect{p}^2}{2m_{\mathrm{ph}}}
    - 
    4 A
  \right)
  = 
  0,
\end{displaymath}
which will exist if and only if $A > 0$.

Physically, this says that the Goldstone mode will be unstable for
$0<|\vect{p}|<p_c$ if the ``static compressibility'',
$\Re\left[K_1^R(0,0)\right]>0$.
In equilibrium, the expression for the component $K_1^R(0,0)$ is real
and negative, but including pumping and decay, there are regions where
solutions of the gap equation, Eq.~(\ref{sp2}) exist but which are
unstable.
Since $\Re\left[K_1^R(0,0)\right]$ is the real part of the second derivative
of the action w.r.t $\psi(\omega=0,k=0)$, it can also be seen as a derivative
of the gap equation, thus unstable solutions are characterised by a non-linear
susceptibility that increases as coherent field increases.

As a result, there are ranges of the parameters $\kappa,\gamma, \mu_B$
for which neither the normal state, nor the ansatz of Eq.~(\ref{ansatz})
are stable.
We have not investigated what alternate stable solutions might exist
under these conditions, however the existence of a real pole in the
response at a non-zero momentum might suggest one should investigate
the possibility of a coherent field at non-zero $\vect{p}$.
Such a possibility would not be too surprising, as spontaneous pattern
formation is seen in laser systems with a continuum of
modes\cite{Denz03}.

\section{Numerical analysis of the mean-field}
\label{sec:numer-analys-mean}

\subsection{Phase diagram}
\label{sec:phase-diagram}

\begin{figure}[htpb]
  \centering
  \includegraphics[width=0.98\columnwidth]{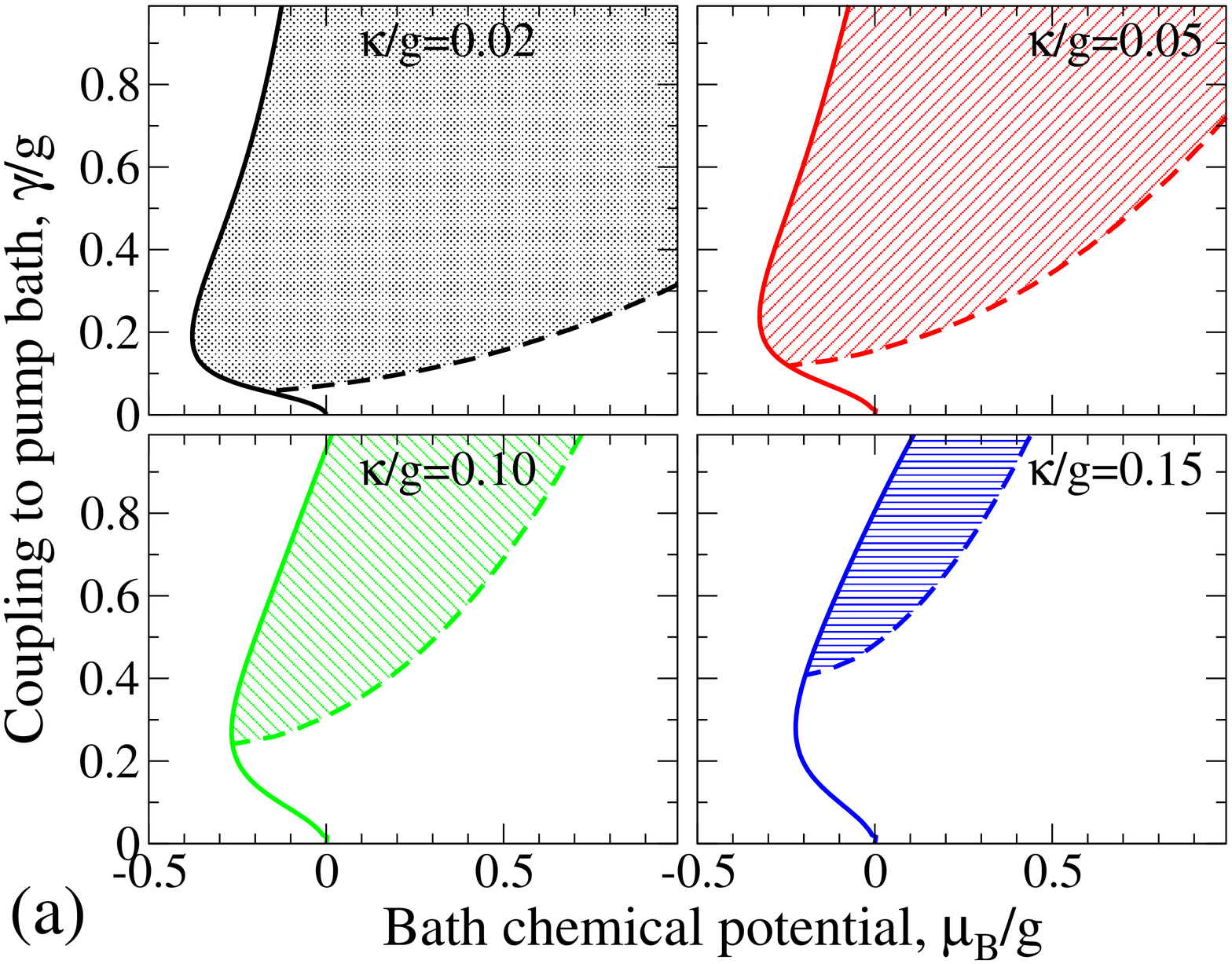}
  \includegraphics[width=0.98\columnwidth]{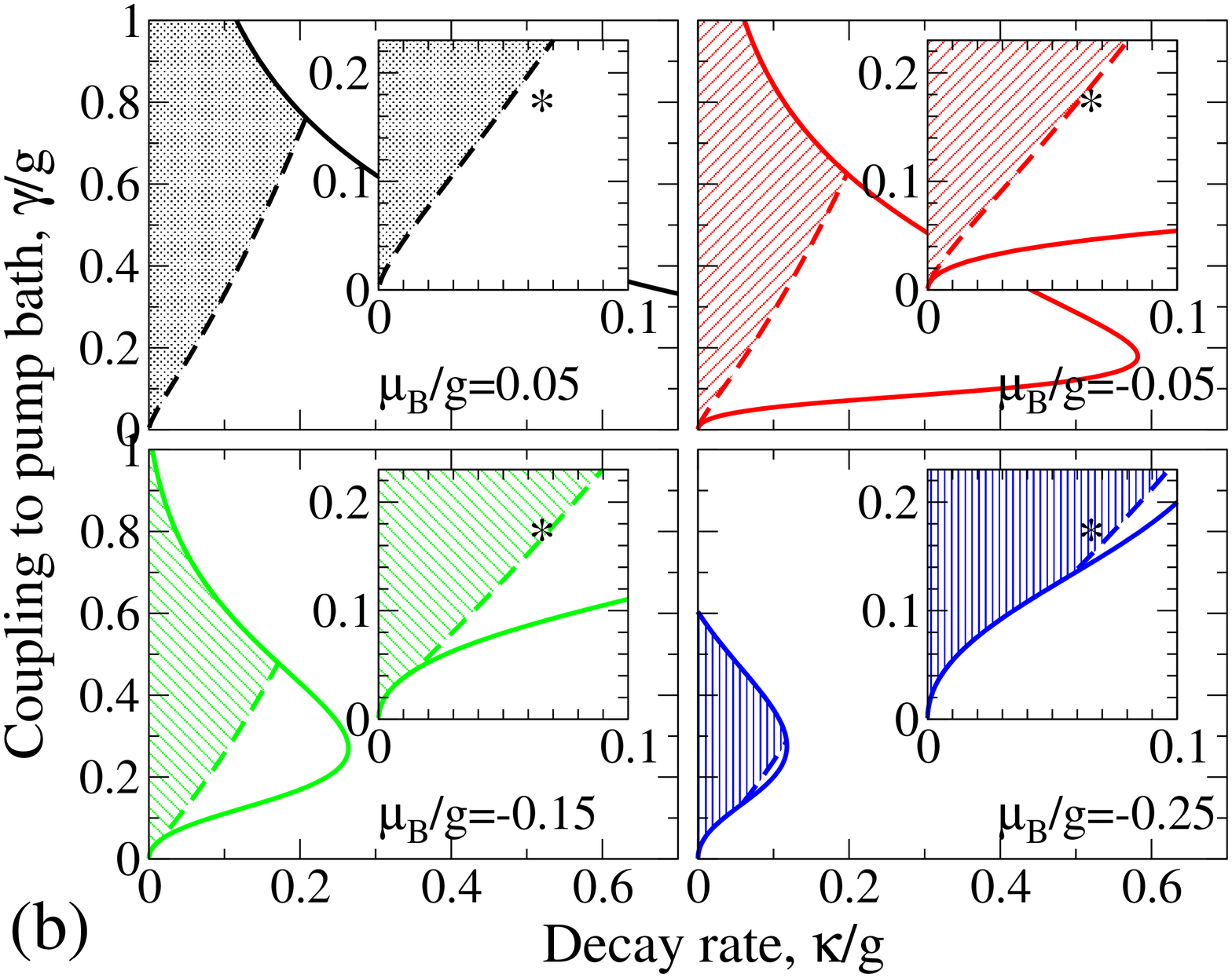}
  \caption{(Color online) Phase boundaries at zero temperature,
      no inhomogeneous broadening.  Panel (a): Fixed decay rate,
      $\kappa$.  Panel (b): Fixed chemical potential, $\mu_B$ (Note
      that $\mu_B > 0$ implies inversion in the pumping bath but not
      necessarily in the system).  The insets show in detail the
      region of small $\kappa$ and $\gamma$.  Solid lines mark the
      limit of stability of the normal state.  Dashed lines mark the
      limit of stability between the uniform condensed state, and some
      other unknown state.  The uniform condensed state is stable in
      the shaded regions.  The asterisk in panel (b) marks a point of
      fixed $\kappa,\gamma$ for comparision between the plots.}  
  \label{fig:kappa-boundary}
\end{figure}

Having discussed the conditions under which the uniform, single-frequency
condensed solution is stable, we may now consider an effective phase boundary
--- i.e. find the ranges of parameters for which there is a stable condensed
solution.
For numerical analysis we choose all baths to be individually in thermal
equilibrium.  However, as the baths need not be in equilibrium with each
other, the system can still be far from thermal equilibrium.
Since the cavity photon modes start at energies much above the zero for bulk
photon modes, we take the chemical potential of the decay bath to be large and
negative.
In addition, since at room temperature the population
of the bulk photon modes at the energy of cavity modes is negligible,
we consider the decay bath to be always at zero temperature.
In the following we will first present calculations at zero pumping
bath temperature, with a delta function density of states i.e.
$\epsilon_{\alpha}=\epsilon$, and at zero detuning.
Following that we will then analyse the influence of finite
temperature of the pumping baths, and of inhomogeneous broadening of
excitons.
At zero bath temperature, the bath distributions are entirely defined
by their chemical potentials, and so there remain three control
parameters, $\mu_B, \gamma, \kappa$.
Note that in this case the pump and decay baths are at the same
temperature, but have very different chemical potentials, thus leading
to a particle flux through the system, driving it out of equilibrium.
In Fig.~\ref{fig:kappa-boundary}, we illustrate the boundary as
a function of $\mu_B, \gamma, \kappa$ by plotting its section in two planes;
the plane of fixed $\kappa$ [Fig.~\ref{fig:kappa-boundary}(a)], and
the plane of fixed $\mu_B$ [Fig.~\ref{fig:kappa-boundary}(b)].

It is worth noting that, for $\mu_B \le 0$, and fixed
$\kappa,\mu_B$ there is both an upper and lower critical $\gamma$.
The maximum $\gamma$ is always present (i.e. even if $\mu_B > 0$), and
results because increased coupling to the bath causes dephasing.
Let us discuss the origin of the minimum critical $\gamma$.
If the bath is at zero temperature, it pumps only that part of the
effective excitonic density of states with energy less than the
  bath chemical potential $\mu_B$.
If there is no inhomogeneous broadening (i.e.
$\epsilon_{\alpha}=\epsilon$) then the effective exciton
density of states is set entirely by its coupling to the baths;
i.e. it is Lorentzian with width $\gamma$.
Thus, the efficiency of pumping depends on how, by broadening the
excitonic energy, the pump leads to a non-zero density of
states below the chemical potential $\mu_B$.
As a result, at $\gamma=0$ there is no pumping, and so no
condensation, and a minimum $\gamma$ is required before there is
sufficient gain to overcome the decay.
If there is inhomogeneous broadening of exciton energies, or 
the pumping baths are at finite temperature, this effect is less
significant, as is seen in Fig.~\ref{fig:T-sigma-modify-boundary}.

From the boundaries of the stable region, it appears that a uniform condensed
stable solution is only possible if $\kappa\le\kappa_0$, with $\kappa_0\simeq
0.2g$. 
The origin of this upper critical $\kappa$ requires further
investigation.

\subsection{Coherent fields and densities}
\label{sec:coher-fields-dens}

\begin{figure}[htpb]
  \centering
  \includegraphics[width=0.98\columnwidth]{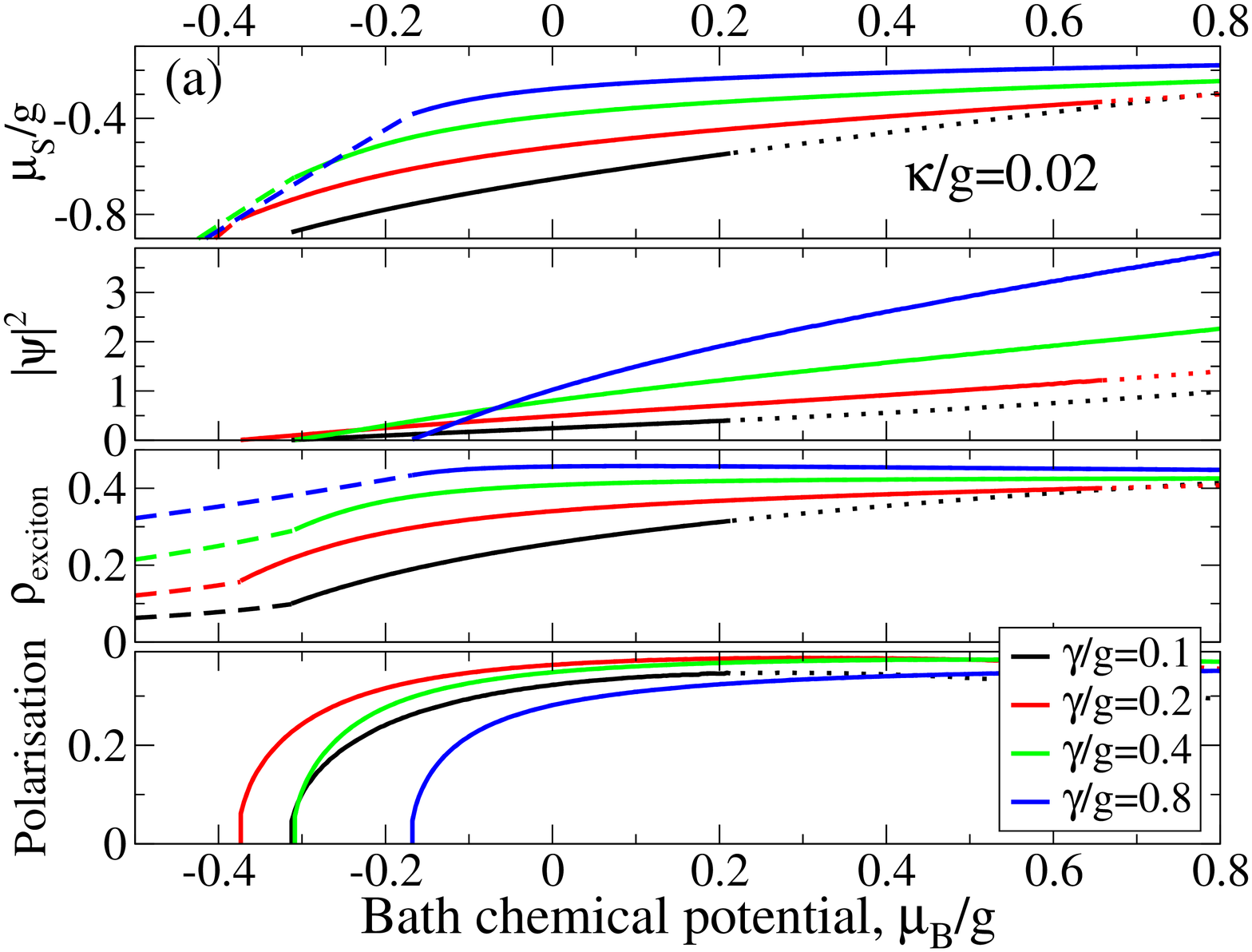}
  \includegraphics[width=0.98\columnwidth]{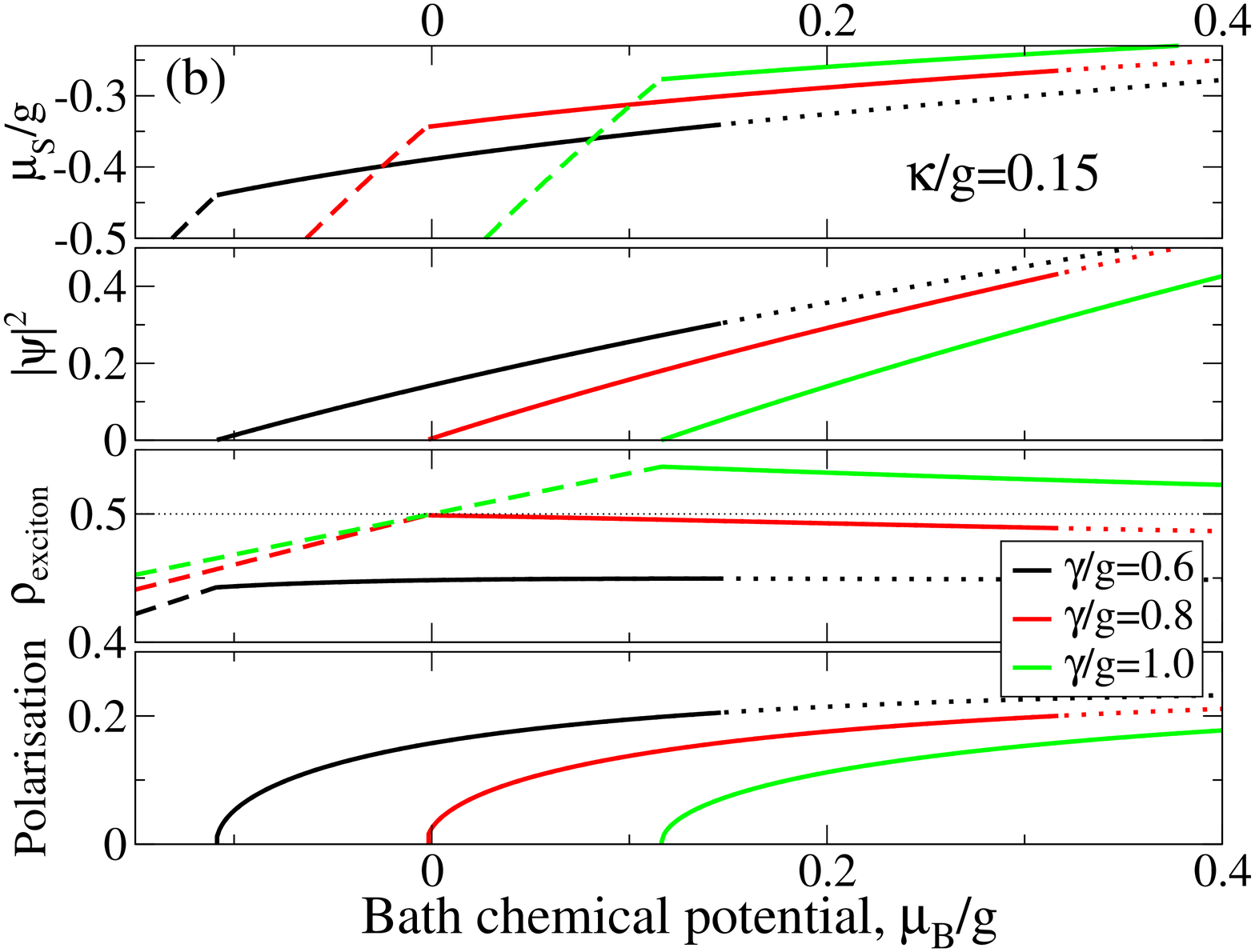}
  \caption{(Color online) Properties of the system as a function of
    bath chemical potential.  Plotted for $T=0$, and
    $\epsilon_{\alpha}=\epsilon$.  Panel (a): $\kappa=0.02g$, panel
    (b) $\kappa=0.15g$, values of $\gamma$ as indicated.  Within each
    panel, four graphs are shown.  Top: The common oscillation
    frequency in the ansatz of Eq.~(\ref{ansatz}), measured from
    $\omega_{\vect{p}=0}=0$; Second: Density of condensed photons;
    Third: exciton density from Eq.~(\ref{eq:rhofermi}); Bottom:
    Polarisation, given by $\sqrt{\tilde{\omega}^2 + \kappa^2}
    \psi_f$.  Solid lines indicate where a stable condensed solution
    exists; dotted lines are the (unphysical) result of the uniform
    condensed solution when it is unstable.  Dashed lines for $\mu_S$
    and $\rho_{\mathrm{exciton}}$ show the comparable quantities in the
    normal state. }
  \label{fig:condensed-trace}
\end{figure}

As well as the phase boundary, one may study the evolution of a number
of properties of the condensate --- e.g. mean-field density of
condensed photons, $|\psi_f|^2$, excitonic density [from
Eq.~(\ref{spd1})], and thus the total mean-field density, being the
sum of condensed photon and exciton densities, polarisation
$\left< a^\dagger b \right>$ (where $|\left< a^\dagger b \right>|^2$
gives the number of condensed fermion pairs - excitons), and
common oscillation frequency $\mu_S$.
These are shown in Fig.~\ref{fig:condensed-trace}, for two values
of $\kappa$ and a range of different $\gamma$, chosen to
illustrate both the regime of weak coupling to baths, where the results
are similar to those in thermal equilibrium, and also strong decay and
pumping, for which the results are instead comparable to the laser.
For comparison, the value of $\mu^{\mathrm{eff}}$, and the
fermion-pair (excitonic) density in the normal state are
shown, which connect smoothly to the condensed quantities, as expected
for a second order phase transition.
Note that for $\kappa=0.15, \gamma=0.9, 1.0$ the excitonic density
$\rho_{\mathrm{exciton}}>0.5$ indicating inversion as is expected
in the lasing case.

\subsection{Influence of bath's temperatures and excitonic density of states}
\label{sec:infl-baths-temp}

\begin{figure}[htpb]
  \centering
  \includegraphics[width=0.98\columnwidth]{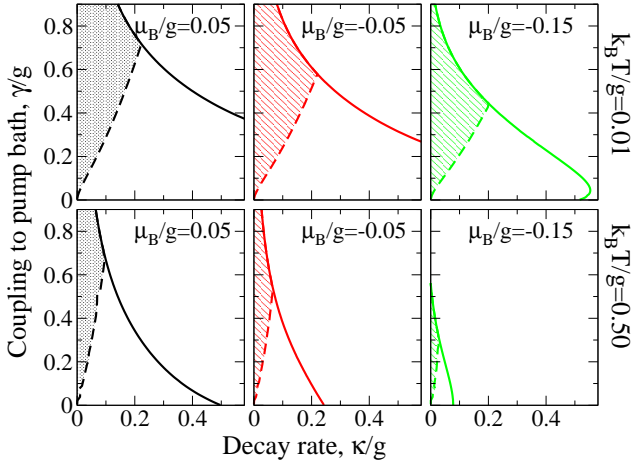}
  \caption{(Color online) Phase boundary for constant chemical
    potential, $\mu_B$, as in Fig.~\ref{fig:kappa-boundary}(b), but
    with a Gaussian distribution of excitonic energies,
    $\sigma_\epsilon=0.15g$, and non-zero temperature (top row
      $T=0.01g$, bottom row $T=0.5g$)}.  As a result, the requirement
    for a minimum coupling strength, $\gamma$, before a transition
    occurs is removed for some phase boundaries.  Solid lines, dashed
    lines, and shaded region mark instability of normal state,
    instability of uniform condensed state, and stable condensed
    region as in Fig.~\ref{fig:kappa-boundary}.
  \label{fig:T-sigma-modify-boundary}
\end{figure}

We now consider the effects of finite bath temperature, and of the
inhomogeneous broadening of the exciton energies.
As such calculations are numerically intensive, we present a limited,
but illustrative set of results.
In Fig.~\ref{fig:T-sigma-modify-boundary}, the equivalent of
Fig.~\ref{fig:kappa-boundary}(b) is shown, but with a Gaussian density of
states, and at small but non-zero temperature of the pumping bath
(the decay bath, of bulk photon modes, is still at T=0).
One can clearly see that by adding inhomogeneous broadening,
$\sigma_{\epsilon}=0.15g$, the lower critical $\gamma$ has been
modified, and for large $\mu_B$ entirely eliminated.
The inset of Fig.~\ref{fig:T-sigma-modify-boundary} shows a higher
temperature, for which none of the curves show any lower critical
$\gamma$.

\begin{figure}[htpb]
  \centering
  \includegraphics[width=0.98\columnwidth]{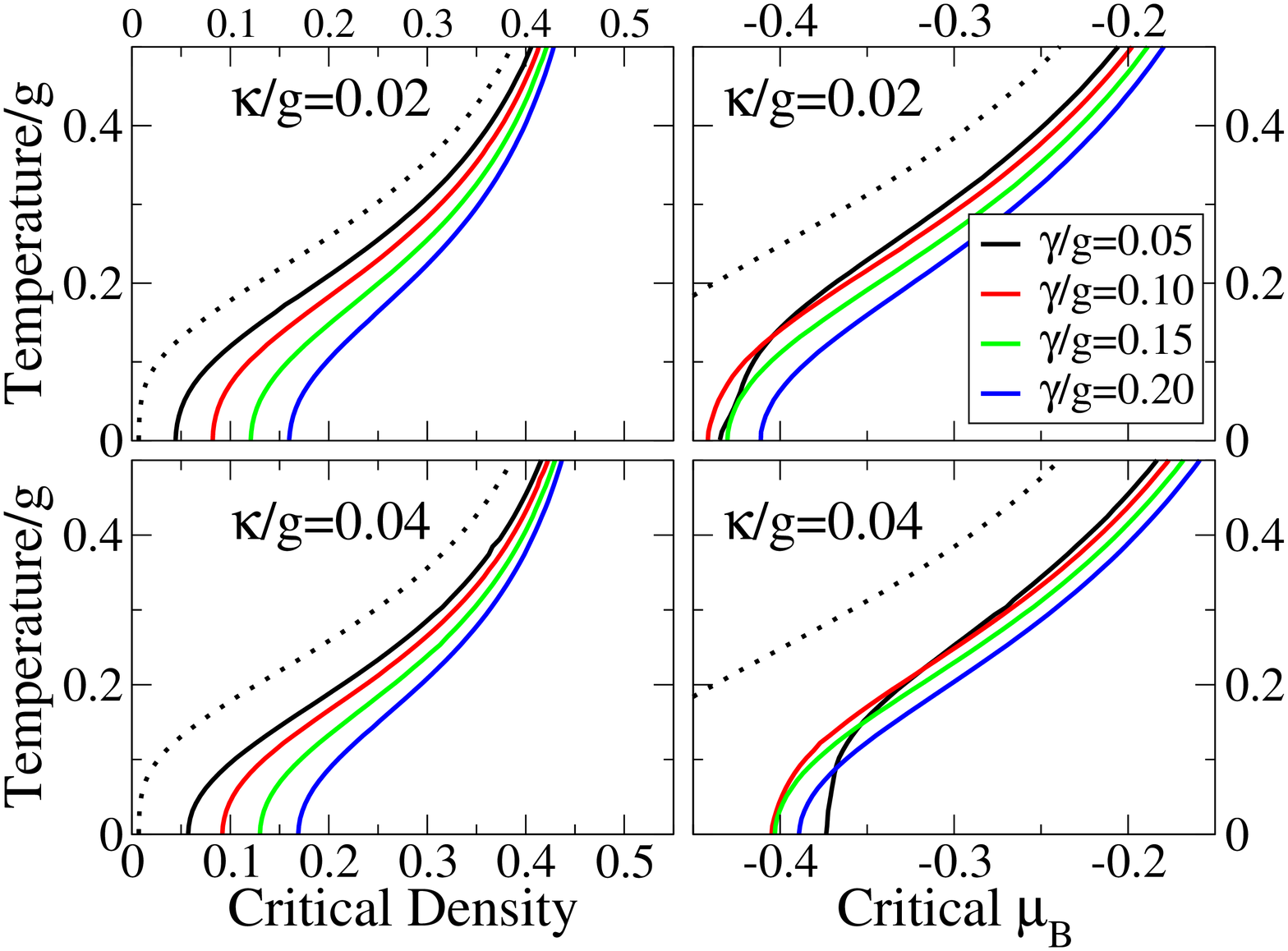}
  \caption{(Color online) Critical density (and associated critical
    bath chemical potential) at a given non-zero bath temperature.
    Evaluated for a Gaussian density of states, with
    $\sigma_{\epsilon}=0.15g$ and values of $\kappa$ and $\gamma$ as
    indicated in the legend. The dotted line marks the limit
    $\kappa\to 0$, $\gamma \to 0$, for which the equilibrium result,
    with distributions set by the pumping bath is recovered.}
  \label{fig:critical-T}
\end{figure}

One can also plot a phase boundary at fixed $\gamma,\kappa$ as a
function of pumping bath temperature $T$ and $\mu_B$, or alternatively
derive the excitonic density $\rho_{\mathrm{exciton}}$ from
Eq.~(\ref{spd1}) to plot the boundary as a function of $T$ and
density.
By doing this we can investigate the influence of decoherence and
particle flux introduced by pumping and decay on the phase diagram,
which can still be significant, even if the system
distribution function would be close to thermal.
For the parameters chosen for the figures, we are in the regime of
densities where the phase transition is well described by mean-field
theory, and so the number of incoherent photons at the transition is
small.
Thus, in this regime, the distribution function of excitons
below and at the transition is set by the pumping bath; thus if the
pumping bath is thermal, then the exciton distribution is too.
This means we can study the influence of dephasing due to pumping and
decay separately from the influence of non-thermal distribution functions.
This also allows direct comparison to the equilibrium limit, which, as
discussed in Sec.~\ref{sec:gamma-=-0} should be recovered as
$\kappa\to 0, \gamma\to 0$.
This is illustrated in Fig.~\ref{fig:critical-T}, where the critical
bath temperature as a function of system density is plotted (and for
comparison, the critical $\mu_B$ at each temperature is also shown).
It is apparent that the presence of pumping and decay shifts the phase
boundary to higher densities, and that the $\rho \to
0$ as $T\to0$ behaviour seen in equilibrium does not survive.
Physically, this increase of critical density is due to the decoherence
introduced by pumping and decay.
The behaviour at $T\to 0$ 
is unsurprising, as the limit $\rho\to 0$ corresponds
to the equilibrium chemical potential $\mu \to -\infty$.
In the presence of non-zero decay rate $\kappa$, one requires a
non-zero effective gain (imaginary part of gap equation), and so no
solution exists with $\mu_B \to -\infty$ even at $T=0$,
 i.e the critical density never goes to zero.

\section{Fluctuations in condensed state to all orders in phase}
\label{sec:fluct-cond-state}

The low energy modes of the broken symmetry system correspond
to slow phase variations.
Since there is no cost to global phase rotations, the action depends
only on derivatives of the phase, and so phase fluctuations may become
large.
Thus, describing $\psi = \psi_0 + \delta \psi$  and considering
only terms to second order in $\delta \psi$ may underestimate how phase
fluctuations reduce long range coherence.
Therefore, we will instead consider the parameterisation $\psi=\sqrt{\rho_0 +
\pi} e^{i \phi}$, and evaluate correlation functions of $\psi$ in terms of the
correlation functions of amplitude $\pi$ and 
phase $\phi$, including the phase fluctuations to all orders.
In equilibrium, the effect of phase fluctuations on the field-field
correlator is responsible for the reduction from long range order to
power law correlations in two dimensions, and so has been much studied
(see e.g. Refs.~\onlinecite{Nagaosa,Popov}).
Here, in order to calculate the luminescence and absorption spectrum, we
will however need also to include density fluctuations.

Combining such a reparameterisation of the fields with the
non-equilibrium Keldysh formalism requires a little care.
The first important consideration is that the parameterisation requires one
to work with fields where $\langle | \psi |^2 \rangle$ is
macroscopic.
This means we should re-parameterise the fields $\psi_f, \psi_b$
defined on the forward and backward contour (see
Sec.~\ref{sec:path-integr-form}), as opposed to the fields $\psi_q,
\psi_{cl}$, since $\langle | \psi_q |^2 \rangle$ is not macroscopic.
This consideration is similar to the fact that the parameterisation
should be done for the fields as functions of space and time
rather than functions of $\vect{p}$ and $\omega$.
The second consideration is that, in calculating the physical
correlation functions, ${\cal D}^{<,>}$, this will involve cross
terms between the two branches, and so one must keep track of which
branch $\pi$ and $\phi$ are on.

The technical details of how to derive the field-field correlation
functions in terms of amplitude and phase Green's functions are
presented in Appendix~\ref{sec:eval-field-corr}.
For the forward Green's function (corresponding to luminescence),
the result is found to be:
\begin{multline}
  \label{eq:final-lum}
  i \mathcal{D}^{<}_{\psi^{\dagger}\psi}(t,r)
  =
  \rho_0 \left\{
    1
    +
    \frac{i}{2\rho_0}
    \left[
      i \mathcal{D}^{<}_{\phi\pi}(t,r)
      -
      i \mathcal{D}^{<}_{\pi\phi}(t,r)
    \right]
  \right.
  \\
  -
  \frac{1}{4\rho_0^2}
  \left[
    i \mathcal{D}^{<}_{\pi\pi}(0,0)
    -
    i \mathcal{D}^{<}_{\pi\pi}(t,r)
    \vphantom{\mathcal{D}^{<}_{\phi}}
  \right]
  \\
  +
  \frac{1}{8\rho_0^2}
  \left[
    i \mathcal{D}^{<}_{\phi\pi}(0,0)
    +
    i \mathcal{D}^{<}_{\pi\phi}(0,0)
    -
    i \mathcal{D}^{<}_{\phi\pi}(t,r)
    -
    i \mathcal{D}^{<}_{\pi\phi}(t,r)
  \right]^2
  \\
  \left.\vphantom{\frac{1}{8}}\right\}
  \exp \left\{
    -
    \left[
      i \mathcal{D}^{<}_{\phi\phi}(0,0)
      -
      i \mathcal{D}^{<}_{\phi\phi}(t,r)
    \right]
  \right\}  
\end{multline}
The above procedure includes amplitude fluctuations $\pi$ and
gradients of phase fluctuations $\nabla\phi,\partial_{t}\phi$ to
second order as they both have restoring force, and cost energy, so
that they are expected to be small.
The phase fluctuations $\phi$ however may be large and in the above
result are taken to all orders.

\begin{figure}[htpb]
  \centering
  \includegraphics[width=0.98\columnwidth]{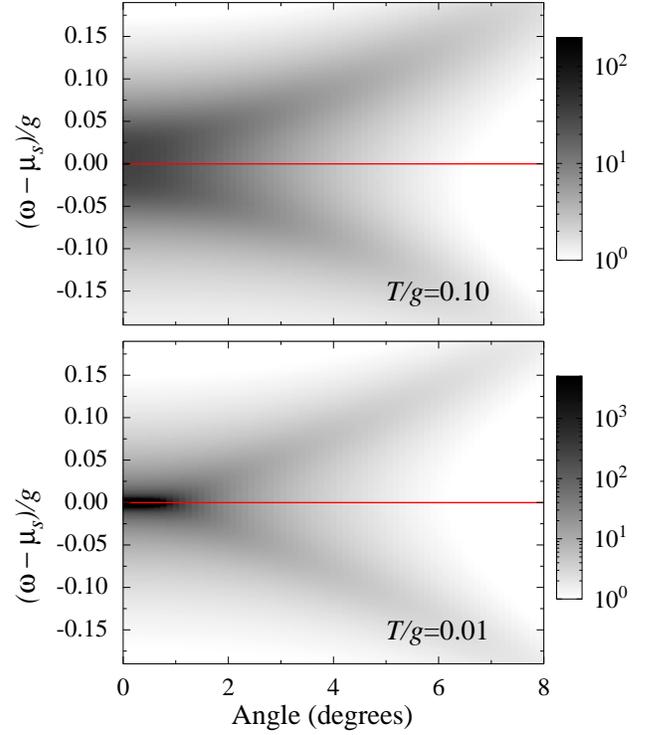}
  \caption{(Color online) Photoluminescence from the region of small
    $\omega$ and small $p$ (shown it terms of an angle of emission
    $\theta = \tan^{-1} (cp/\omega_0)$, calculated including effects
    of phase fluctuations to all orders.  Both panels have
    $\kappa=0.02g, \gamma=0.2g, \mu_B=0.0g$.  The upper panel has
    $T=0.1g$ as in the middle row of Fig.~\ref{fig:spec-lum-abs},
    while the lower has $T=0.01g$, for which the features described in
    the text appear more sharply.}
  \label{fig:lum-detail}
\end{figure}

To calculate the luminescence spectrum one must then Fourier transform
the result ${\cal D}^{<}_{\psi^{\dagger}\psi}(t,r)$ to give the
spectrum in frequency and momentum space.
The first term in the braces in Eq.~(\ref{eq:final-lum})
proportional to $\rho_0$ describes the emission from the condensate
which is now broadened by the exponential term containing the phase
fluctuations.
It is clear that the phase fluctuations determine the condensate
lineshape and the decay of spatial and temporal coherence.
An example of luminescence as given by Eq. (\ref{eq:final-lum})
is shown in Fig.~\ref{fig:lum-detail}.
We will discuss its features in Section \ref{sec:cond-linesh-effects}.

If one were to assume phase fluctuations were small,
then this expression could be expanded to linear order
in Green's functions, and one would find:
\begin{multline*}
  i \mathcal{D}^{<}_{\psi^{\dagger}\psi}(t,r)
  =
  \rho_0\left\{
    1
    -
    \frac{ i \mathcal{D}^{<}_{\pi\pi}(0,0) }{4\rho_0^2}
    -
    i \mathcal{D}^{<}_{\phi\phi}(0,0)
  \right\}
  \\
  +
  \frac{i \mathcal{D}^{<}_{\pi\pi}(t,r)}{4\rho_0}
  +
  \frac{i}{2}\left(
    i \mathcal{D}^{<}_{\phi\pi}(t,r)
    -
    i \mathcal{D}^{<}_{\pi\phi}(t,r)
  \right)
  +
  \rho_0 i \mathcal{D}^{<}_{\phi\phi}(t,r)
\end{multline*}
This is instructive, as the second line describes the fluctuation Green's
function $i \mathcal{D}^{<}_{\delta\psi^{\dagger}\delta\psi}(t,r)$,
 obtained taking the fluctuation fields to second order, while
the first corresponds to a depleted condensate density.
Such a linearisation would describe the luminescence as a sum of two
terms; a condensate term, which due to its lack of space or time
dependence would be a sharp peak, and a fluctuation term.
Furthermore, if one were to consider the frequency spectrum of
fluctuations by integrating this linearised form over momentum one
would have a simple power law form, with a power depending only on the
dimension\cite{Staliunas}, and not on parameters of the system.
\begin{figure*}[htpb]
  \centering
  \includegraphics[width=\textwidth]{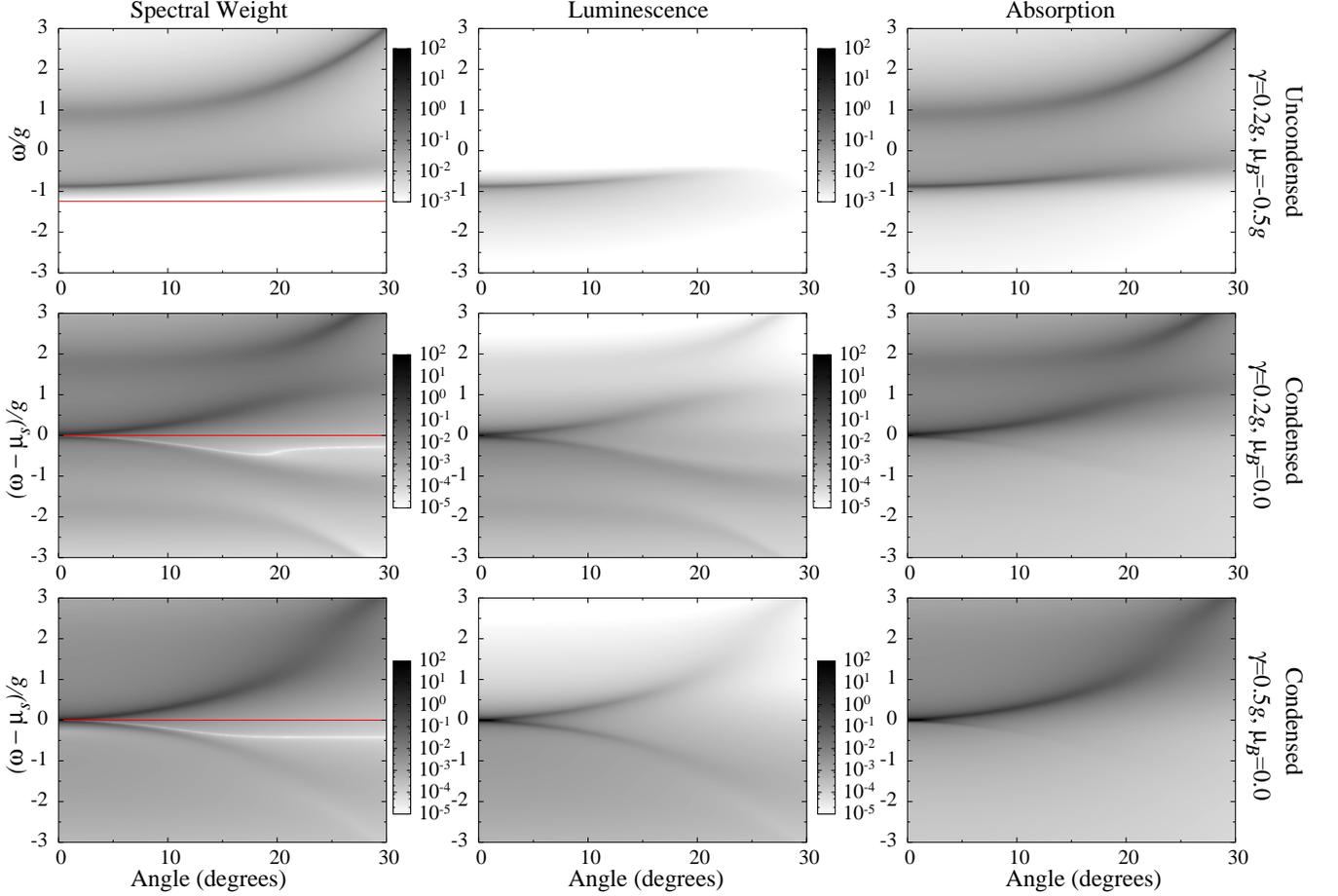}
  \caption{(Color online) Spectral weight, photoluminescence and
    absorption spectra, as a function of emission angle,
    $\tan^{-1}(cp/\omega_0)$.  For all graphs, $\kappa=0.02g$ and
    $T=0.1g$.  Top row: Uncondensed case, $\gamma=0.2g, \mu_B=-0.5g$.
    (cf parameters in Fig.~\ref{fig:normal-luminescence} and
    Fig.~\ref{fig:trace-ns-zeros}) Middle row: Condensed case,
    $\gamma=0.2g, \mu_B=0.0g$.  Bottom row: Condensed case,
    $\gamma=0.5g, \mu_B=0.0g$ (transition to weak coupling).  }
  \label{fig:spec-lum-abs}
\end{figure*}
By allowing phase fluctuations to be large, and keeping the
phase-phase Green's function in the exponent, the condensate acquires
a lineshape as a result of phase fluctuations, and this lineshape can
in the equilibrium limit recover the standard power law correlations
seen in two dimensions.
The form of this lineshape is discussed further in
Sec.~\ref{sec:cond-linesh-effects}.

However, for $\omega, p$ far from zero, such linearisation does not introduce
any major changes; the effects of large phase fluctuations matter mostly at
large times.
Large fluctuations between fields separated by small $t$ or $r$
would imply large gradients, and thus have a large energy cost.
Thus, Fig.~\ref{fig:spec-lum-abs} illustrates the absorption,
luminescence and spectral weight over large ranges of $\omega, p$
using a linearised approach [which at this large scale coincides with
the full expression given by Eq.~(\ref{eq:final-lum})] while
Fig.~\ref{fig:lum-detail}, obtained from the full expression of
  Eq.~(\ref{eq:final-lum}), shows the effect of phase fluctuations at
small $\omega, p$.

For the detailed analysis of the features of the luminescence spectra
we refer to Ref.~\onlinecite{keldysh_letter}. 
Note that for large $\omega, p$ as shown in
Fig.~\ref{fig:spec-lum-abs} the main features of the
non-equilibrium spectra are similar to those predicted for
equilibrium condensation in Refs.~\onlinecite{Keeling,Marchetti}.
In the normal state one can see the upper and lower polariton modes
(top row of Fig.~\ref{fig:spec-lum-abs}) in the spectral weight and
absorption, and only the lower polariton in the luminescence as the
upper polariton is not occupied at this low power.
When system condenses (middle row of Fig.~\ref{fig:spec-lum-abs}) the
structure of modes changes dramatically showing the pairs of phase and
amplitude modes above and below the chemical potential.
Finally, when the coupling to the pump baths; i.e.\ the pumping
  strength is further increased (bottom row of
Fig.~\ref{fig:spec-lum-abs}) the system crosses to weak-coupling
regime and the polariton splitting is suppressed.
In Fig.~\ref{fig:spec-lum-abs} the occupation of the excited states
will not be thermal, in contrast to the analogous figures in
Refs.~\onlinecite{Keeling,Marchetti}, this is however not easy to
observe on these contour plots.
Also since Fig.~\ref{fig:spec-lum-abs} corresponds to pumping baths
at finite temperature in contrast to zero temperature in
Ref.~\onlinecite{keldysh_letter} the sharp occupation edge visible
there is here smeared out.
However the main qualitative difference between the spectra of a
pumped decaying condensate presented here and that for a closed system
given in Refs.~\onlinecite{Keeling,Marchetti} is most visible on small
$\omega, p$ scale as presented in Fig.~\ref{fig:lum-detail}.
This will be discussed in detail in the Section
\ref{sec:cond-linesh-effects}.

\subsection{Condensate Lineshape - effects of dissipation and
  low-dimensionality on decay of correlations}
\label{sec:cond-linesh-effects}

The long range field-field correlations are influenced by the
properties of the soft phase modes; i.e. the Goldstone or Bogoliubov
modes\cite{Wouters05,keldysh_letter}.
By considering the asymptotic behaviour of the phase-phase correlator
at small frequencies and momenta, one can thus find the asymptotic
form of the field-field correlator.
In an equilibrium two-dimensional system, the long distance
field-field correlations decay with a power law below the BKT
transition.
We will now investigate how this asymptotic behaviour is affected by
the presence of pump and decay.
For convenience let us rewrite Eq.~(\ref{eq:two-branch-result}),
assuming an isotropic system:
\begin{align}
  \label{eq:field-correlations-approx}
  i \mathcal{D}^{<}_{\psi^{\dagger}\psi}(t,r) 
  &=
  \rho_0 \left[
    1
    +
    \mathcal{O}\left(1/\rho_0\right)
  \right]
  \exp\left[ -f(t,r) \right],
  \\
  f(t,r)
  \label{eq:correlator-exponent}
  &=
  \int \frac{d\omega}{2\pi}
  \int \frac{p d p}{2\pi}
  \left[
    1-
    J_0(pr) e^{ i\omega t} 
  \right]
  i \mathcal{D}^{<}_{\phi\phi}(\omega,p).
\end{align}
Here $J_0(pr)$ is a Bessel function, from the integration over
azimuthal angle.
We are thus interested in the limits $f(t=0,r\to\infty)$ and
$f(t\to\infty,r=0)$, describing the large distance and long time decay.

For comparison, let us first summarise how this method reproduces the
standard result in the equilibrium case.
In equilibrium, the distribution function is a constant
matrix $F(\omega) = 2n_B(\omega) + 1$, and so:
\begin{align}
  i \mathcal{D}^{<}_{\phi\phi}(\omega,p)
  &= \frac{1}{2}\left(F(\omega) - 1\right) \left(
    i \mathcal{D}^{R}_{\phi\phi}(\omega,p)
    -
    i \mathcal{D}^{A}_{\phi\phi}(\omega,p)
  \right)
  \nonumber\\
  \label{eq:eqbm-luminescence}
  &= n_B(\omega) (-2) \Im\left[ \mathcal{D}^{R}_{\phi\phi}(\omega,p) \right]
\end{align}
For an equilibrium coherent system, the low energy modes will be the linear
Goldstone modes  of the form $\omega = c p$.
By analytic continuation of the imaginary time (Matsubara)
Green's function, one finds:
\begin{align}
  \Im\left[ \mathcal{D}^{R}_{\phi\phi}(\omega,p) \right]
  &=
  \Im\left[ \frac{-C}{(\omega + i 0^+)^2 - c^2 p^2} \right]
  \nonumber\\
  \label{eq:eqbm-spectralweight}
  &=
  -\frac{\pi C}{2 cp}\left( 
    \delta(\omega - cp) -  \delta(\omega + cp)
  \right)
\end{align}
And so, combining Eq.~(\ref{eq:correlator-exponent}),
Eq.~(\ref{eq:eqbm-luminescence}) and Eq.~(\ref{eq:eqbm-spectralweight}) one
finds that the singular contribution to $f(t,r)$ is given by:
\begin{equation}
  \label{eq:singular-eqbm}
  f(t,r) = \frac{C}{2\pi\beta c}
  \int_0^{1/\beta c} \frac{dp}{p}
  \left[
    1 - J_0(pr) \cos(cpt) 
  \right] + \ldots,
\end{equation}
where $\beta$ is inverse temperature, and the effect of the thermal
distribution has been approximated by the upper cutoff of the integral.
The lower cutoff is controlled by how $J_0(pr) \cos(cpt)$
approaches $1$ as $p\to 0$, and thus depend on $r$ and $ct$.
For small $p$, the leading term in the expansion for both $\cos(cpt)$
and $J_0(pr)$ is quadratic, and so the lower cutoff for the integral
is given by $p \simeq 1/\sqrt{r^2 + c^2 t^2}$.
Thus, 
\begin{displaymath}
  f(t,r) \simeq \eta \ln \left(
    \frac{\sqrt{c^2 t^2 + r^2}}{\beta c}
  \right).
\end{displaymath}
Thus, one recovers the standard result, and logarithmic behaviour
of $f(t,r)$ leads to power decay of correlation functions, with
$\eta \propto k_B T / \rho_0$.
One can further use this result to find the form of the peak in the
luminescence spectrum, $\mathcal{L}(\omega,p) \propto (c^2 p^2 +
\omega^2)^{(\eta - 3)/2}$, and the integrated luminescence (i.e.
angular profile\cite{Keeling}) $N(p) \propto p^{\eta - 2}$.

Let us now consider the asymptotic form of the Green's function in
the non-equilibrium case.
We shall first consider the retarded Green's function, as the poles of
this function describe the normal modes; the result of calculating
$\mathcal{D}_{\phi\phi}^<$ will, as discussed later, be to introduce
the population of these modes.
The retarded Green's function, using the notation of
Eq.~(\ref{eq:define-K}) can be written as:
\begin{displaymath}
  i\mathcal{D}^R_{\phi\phi}(\omega,p) = 
  \frac{C}{K_1^R(\omega,p)K_1^R(-\omega,p) - K_2^R(\omega) K_2^R(\omega)}
\end{displaymath}
As discussed in Sec.~\ref{sec:fluct-cond-state-1}, the gap equation
implies that $K_1^R(\omega=0,p=0)=K_2^R(\omega=0)$.
Combining this with the symmetries in Eq.~(\ref{eq:dynamic-symmetries}),
one can show that the most general expression, to quadratic order
in $p,\omega$ 
in the denominator can be written as:
\begin{equation}
  \label{eq:asymptotic-retarded-gf}
  \mathcal{D}_{\phi\phi}^R(\omega,p) \simeq
  \frac{C}{\omega^2 - c^2 p^2 + 2 i \omega x},
\end{equation}
where $C,c$ and $x$ are coefficients to be derived from the
full expressions.
Without pumping and decay, $x=0^+$, and one recovers the equilibrium
result.
With non-zero $x$, the poles of the Green's function,  which
define the low energy modes of the system, have the form
\begin{displaymath}
  \omega = - i x \pm i \sqrt{x^2 - c^2 p^2},
\end{displaymath}
and are thus diffusive, rather than dispersive for $p\le
x/c$\cite{keldysh_letter}.
This can be clearly seen in the luminescence shown in Fig
\ref{fig:lum-detail}:
At low momentum, where the real part of the pole vanishes, but
  the imaginary part does not, the luminescence is dispersionless
  (i.e. flat), but broadened.
Such a form should be generic for broken symmetry in a pumped decaying
system, and indeed the same form has been recently seen in a
related context, of coherently pumped polaritons in photonic
wires, described as an optical parametric oscillator\cite{Wouters05},
as well as in a more generic model\cite{Wouters07}.
This result also shows why it was so important to have solved a
complex gap equation, rather than just adding decay rates to the equilibrium
model.
Adding phenomenological decay rates ``by hand'' would lead to 
a form of the retarded Green's function:
\begin{displaymath}
  \mathcal{D}^{R}_{\phi\phi}(\omega,p) 
  =
   \frac{C}{(\omega + i x)^2 - c^2 p^2} 
\end{displaymath}
Such a form does not describe a system with spontaneously broken
symmetry, as there is no pole at $\omega=0,p=0$, and thus such an
approach misses the appearance of a diffusive mode.

Let us now consider ${\cal D}^{<}_{\phi\phi}$, and thus
the effect of the distribution function.
As was discussed in Sec.~\ref{sec:norm-state-excit}, the distribution
function can be expected to diverge at the energy where the imaginary
part of the denominator of the retarded Green's function vanishes.
This is clear at $\omega=0$ (measured relative to the common oscillation
frequency $\mu_S$), due to the presence of a real pole at $\omega=0, p=0$.
However, this divergence will be exactly canceled by the
vanishing of $\mathcal{D}^R(\omega,p) - \mathcal{D}^A(\omega,p)$
as $\omega \to 0$, since both the divergence and the vanishing
are due to the same imaginary part.
Thus, near $\omega=0$, the asymptotic form of
$\mathcal{D}^{<}_{\phi\phi}(\omega,p)$ is the same as that of
$|\mathcal{D}^{R}_{\phi\phi}(\omega,p)|^2$, i.e.:
\begin{displaymath}
  i\mathcal{D}^<_{\phi\phi}(\omega,p) \simeq 
  \frac{C^2}{(\omega^2 - c^2 p^2)^2 + 4  \omega^2 x^2}.
\end{displaymath}
The effect of the distribution will be to introduce some upper energy cutoff.
Thus, the equivalent of Eq.~(\ref{eq:singular-eqbm}) is:
\begin{equation}
  \label{eq:singular-noneqb}
  f(t,r) 
  =
  \frac{\pi C}{2 c^2 x}
  \int_0^{1/\xi_c} \frac{dp}{p}
  \left[ 1  - J_0(pr) d(p,t) \right]
\end{equation}
where the time dependence is described by:
\begin{multline}
  \label{eq:time-function}
  d(p,t)
  =
  e^{-xt}
  \left[
    \frac{x}{\sqrt{x^2-c^2 p^2}} 
    \sinh\left(\sqrt{x^2-c^2 p^2} t \right)
  \right.
  \\
  \left.
    +
    \vphantom{ \frac{x}{\sqrt{x^2-c^2 p^2}} }
    \cosh\left(\sqrt{x^2-c^2 p^2} t \right)
  \right]
\end{multline}

Equation~(\ref{eq:singular-noneqb}) has a similar interpretation
to Eq.~(\ref{eq:singular-eqbm}), a large $p$ cutoff from the
distribution function and a short distance cutoff set by the
coordinates.
For a thermal distribution function, the upper cutoff
would be given by $1/\xi_c \simeq k_B T/ c$.
Although the photon distribution in the pumped decaying system is not
thermal, if the pumping and decay baths are thermal (as considered
earlier), then the photon distribution will vanish for large enough
energies.
As such, we will write $1/\xi_c \simeq E_{\mathrm{max}}/ c$, where
$E_{\mathrm{max}}$ depends on both pumping and decay, and would
reduce to $k_B T$ in equilibrium.
The result is thus $f(t,r) \simeq \eta^{\prime} \ln (1/ Q \xi_c)$,
where $Q$ is the lower cutoff.
However, the form of the lower cutoff can be different, and
depends on the relative values of $r$, $ct$ and $c/x$.
In the two regions of interest defined at the start
of this section, one finds:
\begin{equation}
  \label{eq:limiting-cutoffs}
  Q =
  \begin{cases}
    \frac{1}{c \sqrt{t/x}} & \text{if $r\simeq 0$, $t \to \infty$,} \\
    \frac{1}{r}            & \text{if $r \to \infty$, $t \simeq 0$.}
  \end{cases}
\end{equation}
Inserting this cutoff, one finds
\begin{equation}
  \label{eq:limiting-decays}
  f(t,r) \simeq
  \begin{cases}
    (\eta^{\prime}/2) \ln (c^2 t/ x \xi_c^2)
    & \text{if $r\simeq 0$, $t \to \infty$,} \\
    \eta^{\prime} \ln (r/  \xi_c)
    & \text{if $r \to \infty$, $t \simeq 0$.}
  \end{cases}  
\end{equation}
Thus, there is still power law decay, but due to pumping and decay the
powers for temporal and spatial decay do not match, and since
$\eta^{\prime}$ may depend on $x$, both power laws will differ from
equilibrium.

Since the long time decay is power law, the lineshape will also have a
power law divergence at low frequency, and as such there is no well
defined condensate linewidth in an infinite system.
In fewer than two dimensions, i.e. in a 1D system\cite{Wouters05}, or a
fully confined system such as a laser with discrete modes, the long
time decay will be exponential, and so a linewidth can be found in
such systems.
The crossover between power law and exponential decay in a large but
finite 2D system is discussed in Sec.~\ref{sec:finite-size-effects}.
In three dimensions, the limit of $f(t,r)$ at large times and
distances is finite (as opposed to divergent as in two, one or zero
dimensions).
As a result, there is phase coherence to arbitrarily large distances,
and so, writing the asymptotic values of $f(t,r)$ as $f_\infty$
there is a contribution to the luminescence that goes like:
\begin{align*}
  i \mathcal{D}^{<}_{\psi^{\dagger}\psi}(\omega,\vect{p})
  &=
   \int dt \int d^3\vect{r} 
  \rho_0 e^{-f_\infty} e^{i\omega t + i \vect{p}\cdot\vect{r}}
  +
  \ldots
  \\
  &=
  \rho_0 e^{-f_\infty} \delta(\omega)\delta^3(\vect{p})
  +
  \ldots,
\end{align*}
i.e., in an infinite homogeneous $3D$ system, there would be a peak at
$\omega=0,\vect{p}=0$, with a peak height given by the condensate
density, which is depleted by phase fluctuations.

\section{Finite size effects}
\label{sec:finite-size-effects}

In the previous section we discussed how the continuum of phase modes
leads, in two dimensions, to logarithmic phase-phase correlation
functions as a function of distance and time.
In this section, we consider how confinement, which leads to a discrete
spectrum of phase modes will modify that result.
In a confined system, there will not be translational invariance, and
so the field-field correlation function will in general depend on
both positions, rather than just on separation.
However, if we are interested in the equal-position, long-time limit,
which is relevant for the lineshape, we can then write:
\begin{displaymath}
  f(t,r,r) = - \sum_n \int \frac{d\omega}{2\pi} 
  \frac{ C |\varphi_n(r)|^2 (1 - e^{i \omega t})}{(\omega^2 - \zeta_n^2)^2 + 4 \omega^2 x^2},
\end{displaymath}
where we have introduced the wavefunction $\varphi_n(r)$ and energy
$\zeta_n$ of the $n^{\mathrm{th}}$ phase mode.
It is clear that if $\varphi_n(r) = e^{i p_n r}$, and $\zeta_n = c p_n$, we
recover the previous result.

Let us now discuss briefly the energy spacing $\Delta$ of phase modes
$\zeta_n$.
Schematically, for a box of size $R$, one has $\Delta = c/R$, i.e. the
sound modes, with discrete momentum spacing.
In contrast, the energy spacing of single particle states in such a box
would be $\delta = 1/2 m R^2$. 
Since the sound velocity increases as condensate density increases,
one can have $\Delta \gg \delta$.
[NB  in a harmonic trap, the Thomas-Fermi radius and the sound
velocity have the same dependence on $\rho_0$, so the phase mode level
spacing is the single particle spacing\cite{Stringari96}.  A harmonic
trap is however a special case in this regard.]

To understand how discrete mode spacing modifies $f(t,r,r)$, let us first
reconsider how the logarithm term arose from the integral. 
Schematically, we had:
\begin{displaymath}
  f(t,r,r) \simeq \int_0^Q \frac{dp}{Q} + \int_Q^{1/\xi_c} \frac{dp}{p}
  = \frac{Q}{Q} + \ln\left(\frac{1}{\xi_cQ} \right)
\end{displaymath}
i.e., the dependence on the coordinates, via the cutoff $Q$ is logarithmic,
as the contribution from $p\le Q$ is constant.
For the discrete sum, after integrating over $\omega$, instead of
Eq.~(\ref{eq:singular-noneqb}) we have:
\begin{equation}
  \label{eq:singular-noneqb-finite}
  f(t,r,r) 
  =
  \frac{\pi C}{2x}
  \sum_n^{N} 
  \frac{|\varphi_n(r)|^2}{\zeta_n^2}
  \left[ 1 - d\left(p=\frac{\zeta_n}{c},t\right) \right]
\end{equation}
with $d(p,t)$ as in Eq.~(\ref{eq:time-function}).
The upper cutoff is introduced here by truncating the
sum at $N$ such that $\zeta_N = c/\xi_c \simeq E_{\mathrm{max}}$.
Considering the long time limit, this sum can also be split into two
parts; for modes $\zeta_n \ll 1\sqrt{x/t}$ the summand is effective
energy independent, while for $\zeta_n \gg \sqrt{x/t}$, with the
density of states in 2D, one recovers a log divergence.
However, the existence of these two parts depends on the relative
values of the energy of the lower cutoff $\sqrt{x/t}$, the upper
cutoff $E_{\mathrm{max}}$, and the level spacing $\Delta$.
We assume $E_{\mathrm{max}} \gg \sqrt{x/t}$, which just means considering long
enough time delays, and so there are three important cases:
\begin{enumerate}
\item $\Delta \ll \sqrt{x/t} \ll E_{\mathrm{max}}$.  In this case there are many
  terms contributing to both the small and large $\zeta_n$ sums, and
  so the the result is as for the integral: schematically $f(t,r,r)=1
  + \ln(E_{\mathrm{max}} \sqrt{t/x})$, and there are power laws, as in the
  infinite system.  This case cannot however persist to arbitrarily
  large times.
\item $\sqrt{x/t} \le \Delta \ll E_{\mathrm{max}}$.  At long enough times, the
  previous case will switch to this case.  Here, there are only a few
  terms in the low energy contribution.  A characteristic term, for
  $\zeta_n \ll x$ gives $d(\zeta_n/c,t) \simeq 1- \zeta_n^2 t / 2 x$.
  Since the number of low energy modes is now of order $1$, rather
  than of order 
  $x/t \Delta^2$, the contribution from these modes
  is of order $t/x$, and not of order 1. Thus, the dominant
  contribution is $ f(t,r,r) \simeq (\pi C/2 x) (t/2x)$,  and so the
  decay of field-field correlations is exponential as in a single mode case.
\item $\sqrt{x/t} \ll E_{\mathrm{max}} \ll \Delta$.  In this case, no phase
  fluctuations are populated, i.e. no terms survive in the sum, and so
  the entire system is coherent.  Using $\Delta = c/ R$, this
  condition is equivalently $R \ll \xi_c = c / E_{\mathrm{max}}$, i.e.  the
  ``thermal length'' is larger than the system size\cite{Petrov00}.
\end{enumerate}

To summarise, if temperature is low enough (or in the case of
non-thermal distribution the relevant energy to which the modes are
occupied is small enough), phase fluctuations are frozen out, as one
expects.
If phase fluctuations are not frozen out, there are two limits; at
long enough times, one always sees linear growth of fluctuations,
resulting in the exponential decay of field-field correlations, and
recovery of the standard laser lineshape\cite{Haken:Laser}.
However, for large enough systems, so level spacing is small, there is
a range of time delays during which the growth of phase fluctuations
is logarithmic in time, giving rise to a power-law decay of
field-field correlations, as one would expect in the infinite system.

\subsection{Self-phase modulation}
\label{sec:comp-laser-linew}

The analysis so far shows how, due to finite size, the power law
correlations associated with a continuum of modes change to the
exponential decay of correlations associated with phase diffusion of a
single mode.
There has been previous work on extending the picture of phase
diffusion of a single mode due to pumping noise\cite{Haken:Laser} to the
case of interacting systems, for which there is an additional source
of noise from self-phase modulation
(SPM)\cite{Holland96,Tassone00,Porras03}.
These works suggest that the phase decay rate can be written as $x
\simeq (\Gamma_0 + \rho_0^2 X_{\text{spm}})/\rho_0$, where $\Gamma_0$
is the noise due to pumping, $\rho_0$ the condensate density, and
$X_{\text{spm}}$ proportional to interaction strength.
We wish here to comment briefly on the origin of the SPM term, and how
it may be modified in the case of many interacting modes, with respect
to the case of phase diffusion of a single mode.

The presence of a SPM term can be understood by considering the
evolution of a coherent state, $e^{\sqrt{\rho_0} \psi^{\dagger}}|0\rangle =
\sum_n (\sqrt{\rho^n/n!}) |n\rangle$.
For an interacting single mode system, the number states are eigenstates,
and have energies like $E_n = a n + b n^2$, thus different
number states evolve at different frequencies, and mutually dephase,
leading to a dephasing rate $x_{\text{spm}} \simeq b \rho_0$.
Thus, SPM occurs because number states, not coherent states are
eigenstates of the single mode Hamiltonian.
The eigenstates of the many mode system, including coherent
interactions between modes, such as $\psi^{\dagger}_0 \psi^{\dagger}_0
\psi^{\nodagger}_p \psi^{\nodagger}_{-p}$ are neither number states
nor coherent states, but are instead better described by
Noiz\'erres-Bogoliubov states\cite{Nozieres82}.
Such states are superpositions of terms with different divisions of
particles between the condensate and non-condensed modes; while they
may be eigenstates of total number, they are not eigenstates of the
number of particles in a given mode, and they lower energy because of
the coherence between the different modes\cite{Littlewood04}.
As such, when considering systems with a continuum of interacting
modes, it is not clear that SPM terms should exist, or if they exist,
should have the same form.

\section{Conclusions}
\label{sec:conclusions}
In conclusion, we have studied steady-state spontaneous quantum
condensation in a non-equilibrium Bose-Fermi system with pumping and
decay, and consequent flux of particles.
In order to study the effect of large phase fluctuations in the broken
symmetry system, it was necessary to extend the path-integral Keldysh
formalism to deal with a reparameterisation in terms of phase and
amplitude fluctuations, for fields on the forward and backward time
contours.
We have shown that the mean-field properties of a pumped and
decaying condensate can be described by a complex analogue of the
Gross-Pitaevskii equation in the BEC regime (or equivalently the
gap equation in the BCS regime).
The real part of this self-consistency equation relates the coherent
field to the system's non-linear susceptibility as in the case of
equilibrium condensation, while the imaginary part reflects how the
gain and decay are balanced, as in a laser.
We further show that it is crucial to satisfy this complex
self-consistent equation in order to get the correct collective mode
structure, reflecting the broken symmetry.

We have analysed the solutions of this complex gap equation and
examined their stability.
Surprisingly, despite non-thermal distributions, the instability
of the normal state is analogous to that in thermal equilibrium, 
where the normal state becomes unstable when the chemical potential,
at which the Bose-Einstein distribution diverges, reaches the
bottom of the system's spectrum.
In the non-equilibrium case, the system's distribution,
although far from thermal, develops a divergence at some energy.
When, by tuning parameters of the system, this energy is brought to
coincide with an effective pole of the system's Green's function, then
the normal state becomes dynamically unstable and the condensation
transition takes place.
We have also shown that whenever there is a condensed solution, the
normal state becomes dynamically unstable, and so there is no ambiguity
as to which state the system would choose.
However, we have found a range of parameters where both the normal and
the uniform harmonic condensed solutions are unstable, suggesting
either more exotic, perhaps chaotic, dynamics or spatial pattern
formation.

We have analysed the non-equilibrium phase diagram as a function of
the decay and pump parameters, and have found both the low density
condensed solutions when pump and decay strengths are relatively
small, as well as the high density, inverted, laser-like solutions
when pump and decay are comparable to the inter particle interactions.
When applied to microcavity polaritons these regimes reflect the
spontaneous condensation of strongly coupled photon-exciton modes at
relatively small pump and decay powers, and the crossover to the
weak-coupling regime and the photon laser at large pump powers.
It is important to stress that even if the system distribution is
close to thermal the presence of pump and decay, i.e.\ particle flux,
result in a higher critical density at a given temperature than in a
closed system, and there is a non-zero minimum critical density
even at zero temperature.

Having analysed the fluctuation spectra and collective modes, we have
found an important difference between condensation in a dissipative
environment and that in closed systems: Although there is a real pole
(undamped mode) at zero frequency and momentum, indicating broken
symmetry, the usual linear dispersion of the sound mode (Bogoliubov,
Goldstone mode) at small momenta is now replaced by diffusive
behaviour (i.e.\ a broadened but flat dispersion); this questions the
possibility of superfluidity on large time and distance scales.
This qualitatively new structure of the collective modes is visible in
the luminescence and absorption spectra, and it affects the field-field
correlations, i.e.\ decay of spatial and temporal coherence, and the
condensate lineshape.
For example, in the 2D system dissipation changes the usual power-law
decay of spatial and temporal coherence, replacing it by one where the
powers for temporal and spatial decay do not match.

It is instructive to place our treatment of non-equilibrium
  quantum condensation in the context of other works on dynamic effects
  in polariton systems.  
  Much of the literature concentrates on Boltzmann-like rate equations.
  Such an approach allows one to study the effect of pumping and decay
  on the occupation of modes;\cite{Tassone97,Porras02,savaona06} but
  is not able to account for the changes to the excitation spectrum
  and the density of states (which are particularly dramatic as the
  system crosses the phase transition) due to the pumping, decay and
  presence of the coherent field.
  In contrast, field theoretical studies presented here
  selfconsistently account both for arbitrarily large changes to
  the excitation spectrum as well as changes to the occupation of this
  spectrum.
  Such approaches are thus well placed to study the phase transition
  between the non-condensed and condensed states, and in addition the
  crossover between strong and weak coupling regimes.
  A closer approach to the field-theoretical approach presented here
  would be the evolution of the off diagonal parts of the full density
  matrix\cite{Porras03,Laussy04}.
  It was however only recently that qualitative changes to the
  spectrum have been calculated using the density matrix approach (in
  the context of parametric emission from photonic wires) in
  Refs.~\onlinecite{Wouters05,Wouters06}.
  A further distinction is between single-mode models, in which one
  expects phase diffusion (e.g.
  Refs.~\onlinecite{Porras02,Tassone00}) and exponential decay of
  correlations as in lasers, and models with a continuum of modes,
  such as Refs.~\onlinecite{keldysh_letter,Wouters05,Wouters06} and
  this paper.

Finally, in this paper we have analysed how the finite size of
the system affects the decay of temporal coherence.
This is particularly important for the understanding of recent
experiments\cite{Kasprzak06}, as well as for providing a connection to
similar analysis for single mode photon lasers, which are still used
as the basis to describe the decay of coherence in atom and polariton
lasers.
The key difference between the output from the condensate and from a
single mode laser is that in the condensate there is a continuum of
modes, and so spatial fluctuations play an important role --- in 2D
they destroy the long-range order and lead to a power-law decay of
correlations.
Including such spatial fluctuations, the growth of phase
fluctuations as a function of time is logarithmic, which gives
power-law decay of temporal coherence, rather than the exponential
decay expected for a single mode.
In single mode systems such as the laser, there are no spatial
fluctuations, and so the decay of coherence is determined entirely
from the phase diffusion of this single mode.
However, if one takes a continuum system, and reduces its size, the
energy spacing of modes becomes larger, and so the number of modes
whose energies are low enough to be relevant decreases, eventually
recovering the single mode limit.
We have identified two regimes in the finite system: 
Where the level spacing is larger than temperature, and so spatial
fluctuations are essentially frozen out, resulting in an exponential
decay of correlations as in a single mode laser; and where the level
spacing is small with respect to temperature, so one gets a power-law
decay of temporal coherence at short times as in the infinite system,
crossing over to exponential decay at larger times.

The qualitative implications of our results are general, and can apply
to any BEC or BCS condensate which is subject to dissipation.
The immediate applications of this analysis are for polariton BEC,
which are naturally faced with significant pumping and decay
processes.
However, the techniques and results developed here, can be of use in
understanding a wider class of broken symmetry dissipative systems;
for example resonant parametric oscillators, and atom lasers, where
coherence, dephasing, and the interaction of many modes are all
relevant.

\begin{acknowledgments}
  We are grateful to Ben Simons and Roland Zimmermann for suggestions and
  useful discussions. M.H.S would like to acknowledge stimulating visit to
  Physics Department, Humboldt University, Berlin.  We acknowledge financial
  support from EPSRC (M.H.S.)  and the Lindemann Trust (J.K).
\end{acknowledgments}

\appendix

\section{Gap equation at T=0}
\label{sec:gap-equation-at}

In the limit of $T=0$, the integrals in the gap equation,
Eq.~(\ref{sp2}) can be evaluated in terms of elementary
functions.
This makes numerical analysis of the equations much easier
in this limit.  
The results of this analysis are presented in
Sec.~\ref{sec:numer-analys-mean}; for completeness we show the
explicit expressions at $T=0$ here.
At $T=0$, the bath distributions take a simple form
$F_b(\omega)=\rm{sign}(\omega-\tilde{\mu}_B)$ and
$F_a(\omega)=\rm{sign}(\omega+\tilde{\mu}_B)$ and so the real and the
imaginary parts of the gap equation become:
\begin{multline*}
  \tilde{\omega}_0
  = 
  - \frac{g^2\gamma \tilde{\epsilon}}{2\pi2E(E^2+\gamma^2)}
  \rm{ln}
  \frac{%
    (E+\mu_B-\frac{\mu_S}{2})^2 + \gamma^2
  }{%
    (E-\mu_B+\frac{\mu_S}{2})^2+\gamma^2
  }
  + \\
  \frac{g^2}{2\pi E}\left(
    \rm{ArcTan}\frac{E+\mu_B-\frac{\mu_S}{2}}{\gamma}+
    \rm{ArcTan}\frac{E-\mu_B+\frac{\mu_S}{2}}{\gamma} 
  \right) 
  - \\
  \frac{g^2\tilde{\epsilon}}{2\pi(E^2+\gamma^2)}
  \left(
    \rm{ArcTan}\frac{E+\mu_B-\frac{\mu_S}{2}}{\gamma}
  \right.
  - \\
  \left.
    \rm{ArcTan}\frac{E-\mu_B+\frac{\mu_S}{2}}{\gamma} 
  \right),
\end{multline*}
and
\begin{multline*}
  \frac{\kappa}{\gamma}
  = 
  \frac{g^2\gamma}{2\pi2E(E^2+\gamma^2)}
  \rm{ln}
  \frac{%
    (E+\mu_B-\frac{\mu_S}{2})^2 + \gamma^2
  }{%
    (E-\mu_B+\frac{\mu_S}{2})^2+\gamma^2
  } 
  + \\
  \frac{g^2}{2\pi(E^2+\gamma^2)}
  \left(
    \rm{ArcTan}\frac{E+\mu_B-\frac{\mu_S}{2}}{\gamma}
  \right.
  - \\
  \left.
    \rm{ArcTan}\frac{E-\mu_B+\frac{\mu_S}{2}}{\gamma} 
    \right).
\end{multline*}
The expression for fermion-pair (exciton) density~(\ref{spd1})
at $T=0$ reduces to:
\begin{multline*}
  \frac{1}{2}\left(b^{\dagger}b-a^{\dagger}a\right) 
  =
  \frac{g^2\gamma |\psi_f|^2}{4\pi E(E^2+\gamma^2)}
  \rm{ln}\frac{%
    (E-\mu_B+\frac{\mu_S}{2})^2 + \gamma^2
  }{%
    (E+\mu_B-\frac{\mu_S}{2})^2 + \gamma^2
  } 
  \\
  -\frac{\tilde{\epsilon}}{2\pi E}
  \left(
    \rm{ArcTan}\frac{E-\mu_B+\frac{\mu_S}{2}}{\gamma} 
    +
    \rm{ArcTan}\frac{E+\mu_B-\frac{\mu_S}{2}}{\gamma} 
  \right) 
 \\
 +
  \left(
    \frac{g^2 |\psi_f|^2}{2\pi(E^2+\gamma^2)}-\frac{1}{2\pi}
  \right)
  \left(
    \rm{ArcTan}\frac{E-\mu_B+\frac{\mu_S}{2}}{\gamma}
  \right.
  - \\
  \left.
    \rm{ArcTan}\frac{E+\mu_B-\frac{\mu_S}{2}}{\gamma} 
  \right).
\end{multline*}

\section{Evaluation of field correlations in terms of amplitude and phase fluctuations}
\label{sec:eval-field-corr}

To illustrate the idea of using phase and amplitude fluctuations,
  we will first present the simpler case of ${\cal D}^{T,\tilde{T}}$,
  for which both fields are on the same branch, and so we may drop all
  labels identifying which branch or Green's function ($T$ or
  $\tilde{T}$) we are considering.
Then, writing $\pm$ for the time, and coordinate indices $(T\pm t/2,
\vect{R}\pm\vect{r}/2)$, one may write:
\begin{align*}
  i{\mathcal{D}}_{\psi^{\dagger}\psi}
  &=
  \left< \psi^{\dagger}(+) \psi^{\nodagger}(-) \right> 
  \\
  &=
  \left< 
    \sqrt{(\rho_0 + \pi(+))(\rho_0 + \pi(-))}
    e^{-i(\phi(+)-\phi(-))}
  \right>
\end{align*}
The square root may be expanded to second order in the density
fluctuations (as density fluctuations, unlike phase fluctuations, have
a restoring force), thus:
\begin{widetext}
  \begin{displaymath}
    i{\mathcal{D}}_{\psi^{\dagger}\psi}
    \simeq
    \rho_0\left< 
      \left[
        1 
        + 
        \left( \frac{\pi(+) + \pi(-)}{2\rho_0} \right)
        - 
        \frac{\left(\pi(+) - \pi(-)\right)^2}{8\rho_0^2}
      \right]
      \exp\left[-i(\phi(+)-\phi(-))\right]
    \right>.
  \end{displaymath}
  Introducing a current $J$, one may write the correlators in
  terms of a generating functional as:
  \begin{multline*}
    i{\mathcal{D}}_{\psi^{\dagger}\psi}
    =
    \rho_0
    \left\{
      1 
      + 
      \sum_{\omega,p}
      \frac{1}{\rho_0} 
      \cos\left(
        \frac{\omega t}{2} + \frac{\vect{p}\cdot\vect{r}}{2}
      \right) 
      \frac{\delta}{\delta J_{\omega,p}}
      +
      \frac{1}{2\rho_0^2}
      \left[
        \sum_{\omega,p}
        \sin\left(
          \frac{\omega t}{2} + \frac{\vect{p}\cdot\vect{r}}{2}
        \right) 
        \frac{\delta}{\delta J_{\omega,p}}
      \right]^2
    \right\}
    \nonumber\\
    \times
    \left.
      \left< 
        \exp\left[
          \sum_{\omega,p}
          J_{\omega,p} \pi(\omega,p) +
          2 \sin\left(
            \frac{\omega t}{2} + \frac{\vect{p}\cdot\vect{r}}{2}
          \right) 
          \phi(\omega,p)
        \right]
      \right>
    \right|_{J=0}.
  \end{multline*}
\end{widetext}
By integrating over the photon field, the generating functional, ${\cal Z}[J]
= \langle \exp[\dots]\rangle$, can be expressed in terms
of the correlators of amplitude and phase fluctuations.
Defining
\begin{displaymath}
  \vect{J}(\omega,p)
  =
  \left(
    \begin{array}{c}
      J_{\omega,p}
      \\
      2 \sin\left[
        \left( {\omega t} + {\vect{p}\cdot\vect{r}} \right)/2
      \right]
    \end{array}
  \right)
\end{displaymath}
one may then write
\begin{equation}
  \label{eq:generating-functional}
  {\cal Z}[J]
  =
  \exp \left[
    \frac{1}{2}
    \sum_{\omega,p}
    \vect{J}(-\omega,p)^{T}
    i \tilde{\mathcal{D}}(\omega,p)
    \vect{J}(\omega,p)
  \right]
\end{equation}
where
\begin{displaymath}
  \tilde{{\mathcal{D}}}
  =
  \left(
    \begin{array}{ll}
      {\mathcal{D}}_{\pi\pi} &
      {\mathcal{D}}_{\pi\phi} \\
      {\mathcal{D}}_{\phi\pi} &
      {\mathcal{D}}_{\phi\phi} 
    \end{array}
  \right).
\end{displaymath}
Note that we use the standard definition of Green's functions so that
$i{\mathcal{D}}_{ab} = \langle a b \rangle$.

To determine these correlators one may either recalculate the
effective action  by writing $\psi$ in terms of $\pi$ and $\phi$ and
expanding to second order in $\pi$ and derivatives of $\phi$; or
one may use the fact that at second order, the amplitude-phase
variables can be considered as a linear transform of $\delta \psi$ and
$\delta \bar{\psi}$, i.e.:
\begin{displaymath}
  \left(
    \begin{array}{l}
      \pi \\ \phi
    \end{array}
  \right)
  =
  L
  \left(
    \begin{array}{l}
      \delta \psi^{\nodagger} \\ \delta \psi^{\dagger}
    \end{array} 
  \right)
  ,
  \quad
  L
  =
  \frac{1}{2 \sqrt{\rho_0}}
  \left(
    \begin{array}{cc}
      2\rho_0 & 2\rho_0 \\
      -i & i
    \end{array}
  \right).
\end{displaymath}
Note that this rotation relates the effective action expressed in
terms of these variables, and is not to be used in finding the
final correlation functions $\mathcal{D}_{\psi^{\dagger}\psi}$.
Thus, one can express the amplitude-phase Green's functions
in terms of the $\delta \psi, \delta \psi^{\dagger}$ Green's
functions as:
\begin{equation}
  \label{eq:amp-phase-rotation}
  \tilde{\mathcal{D}}^{R/A/K} 
  =
  L
  \mathcal{D}^{R/A/K} 
  L^{\dagger}.
\end{equation}
The $T,\tilde{T}$ Green's functions can then be found from the
retarded, advanced and Keldysh components by using
Eq.~(\ref{eq:inv-keldysh-1}) and:
\begin{equation}
  \label{eq:inv-keldysh-2}
  {\mathcal{D}}^{T,\tilde{T}} = \frac{1}{2}
  \left(
    {\mathcal{D}}^{K} 
    \pm \left[
      {\mathcal{D}}^{R} + {\mathcal{D}}^{A}
    \right]
  \right).
\end{equation}
Thus, one may write the $T$ or $\tilde{T}$ field correlation function
in terms of the phase and amplitude Green's functions:
\begin{widetext}
  \begin{multline}
    \label{eq:single-branch-result}
    i{\mathcal{D}}_{\psi^{\dagger}\psi}
    =
    \rho_0
    \left\{
      \vphantom{ \left( \sum_{\omega,p} \frac{1}{2} \right) }
      1 
      -
      \sum_{\omega,p}
      \sin\left(
        \omega t + \vect{p}\cdot\vect{r}
      \right) 
      \frac{i \mathcal{D}_{\phi\pi}(\omega,p)}{\rho_0} 
      -
      \sum_{\omega,p}
      \left[
        1-
        \cos\left(
          \omega t + \vect{p}\cdot\vect{r}
        \right) 
      \right]
      \frac{i \mathcal{D}_{\pi\pi}(\omega,p)}{4\rho_0^2}
    \right.
    \nonumber\\
    \left.
      +
      \frac{1}{2}
      \left(
        \sum_{\omega,p}
        \left[
          1-
          \cos\left(
            \omega t + \vect{p}\cdot\vect{r}
          \right) 
        \right]
        \frac{i \mathcal{D}_{\phi\pi}(\omega,p)}{\rho_0}
      \right)^2
    \right\}
    \exp\left\{
      - \sum_{\omega,p}
      \left[
        1-
        \cos\left(
          \omega t + \vect{p}\cdot\vect{r}
        \right) 
      \right]
      i \mathcal{D}_{\phi\phi}(\omega,p)
    \right\}.
  \end{multline}
\end{widetext}

We can now address how to generalise this calculation when the two
fields are on different branches.
We will consider the forward (luminescence) Green's function;
the backward (absorption) will follow by swapping labels.
Thus, repeating the above discussion, but keeping subscripts 
on the fields one has:
\begin{widetext}
  \begin{displaymath}
    i \mathcal{D}^{<}_{\psi^{\dagger} \psi}(t,r) 
    =
    \rho_0\left< 
      \left[
        1 
        + 
        \left( \frac{\pi_{b}(+) + \pi_{f}(-)}{2\rho_0} \right)
        - 
        \frac{\left(\pi_{b}(+) - \pi_{f}(-)\right)^2}{8\rho_0^2}
      \right]
      \exp\left[-i(\phi_{b}(+)-\phi_{f}(-))\right]
    \right>.
  \end{displaymath}
  Then, as before, introducing a current, we may write this
  in terms of a generating functional.  However, to keep
  track of labels, we shall need two currents, $J_f$ and $J_b$,
  thus:
  \begin{multline*}
    i \mathcal{D}^{<}_{\psi^{\dagger} \psi}(t,r) 
    =
    \rho_0
    \left\{
      1
      +
      \sum_{\omega,p} \frac{1}{2\rho_0} \left[
        \frac{\delta}{\delta J_{b,(\omega,p)}} 
        e^{i(\omega t + \vect{p}\cdot \vect{r})/2} 
        +
        \frac{\delta}{\delta J_{f,(\omega,p)}} 
        e^{-i(\omega t + \vect{p}\cdot \vect{r})/2} 
      \right]
    \right.
    \\
    \left.
      -
      \frac{1}{8\rho_0^2}
      \left(
        \sum_{\omega,p}  \left[
          \frac{\delta}{\delta J_{b,(\omega,p)}} 
          e^{i(\omega t + \vect{p}\cdot \vect{r})/2} 
          -
          \frac{\delta}{\delta J_{f,(\omega,p)}} 
          e^{-i(\omega t + \vect{p}\cdot \vect{r})/2} 
        \right]
      \right)^2
    \right\}
    \left<
      \exp\left(
        \sum_{\omega,p}
        \vect{J}^{T}(\omega,p) \vect{\Lambda}(\omega,p)
      \right)
    \right>.
  \end{multline*}
\end{widetext}

The calculation proceeds as before, but now the generating
functional is written in terms of:
\begin{displaymath}
    \vect{J}(\omega,p) =
  \left(
    \begin{array}{c}
      J_{b,(\omega,p)} \\
      -i e^{+i (\omega t + \vect{p}\cdot\vect{r})/2} \\
      J_{f,(\omega,p)} \\
      i e^{-i (\omega t + \vect{p}\cdot\vect{r})/2}
    \end{array}
  \right)
  , \quad
  \vect{\Lambda}(\omega,p) = 
  \left(
    \begin{array}{c}
      \pi_{b} \\ \phi_{b} \\ \pi_{f} \\ \phi_{f}
    \end{array}
  \right)_{\omega,p}.
\end{displaymath}
There is thus an additional $2\times2$ structure  of the Green's
functions associated with branch labels, i.e. in block notation,
\begin{displaymath}
  \tilde{{\mathcal{D}}}
  =
  \left(
    \begin{array}{ll}
      \tilde{{\mathcal{D}}}^{\tilde{T}} &
      \tilde{{\mathcal{D}}}^{>} \\
      \tilde{{\mathcal{D}}}^{<} &
      \tilde{{\mathcal{D}}}^{T} 
    \end{array}
  \right).
\end{displaymath}

With such an extended matrix structure, one can generalise
Eq.~(\ref{eq:generating-functional}), and thus find the result:
\begin{widetext}
  \begin{multline}
    \label{eq:two-branch-result}
    i \mathcal{D}^{<}_{\psi^{\dagger}\psi}(t,r) 
    =
    \rho_0 \left\{
      1
      +
      \frac{i}{2\rho_0}
      \sum_{\omega,p}
      \left[
        i \mathcal{D}^{T}_{\phi\pi}(\omega,p) - 
        i \mathcal{D}^{\tilde{T}}_{\phi\pi}(\omega,p)
        +
        e^{i(\omega t + \vect{p}\cdot \vect{r})}
        \left(
          i \mathcal{D}^{<}_{\phi\pi}(\omega,p)
          -
          i \mathcal{D}^{<}_{\pi\phi}(\omega,p)
        \right)
      \right]
    \right.
    \\
    \left.
      -
      \frac{1}{4\rho_0^2}
      \sum_{\omega,p}
         \left(
           1 - e^{i(\omega t + \vect{p}\cdot \vect{r})}
         \right)
         i \mathcal{D}^{<}_{\pi\pi}(\omega,p)
      +
      \frac{1}{8\rho_0^2}
      \left[
        \sum_{\omega,p}
         \left(
           1 - e^{i(\omega t + \vect{p}\cdot \vect{r})}
         \right)
        \left(
          i \mathcal{D}^{<}_{\phi\pi}(\omega,p)
          +
          i \mathcal{D}^{<}_{\pi\phi}(\omega,p)
        \right)
      \right]^2
    \right\}
    \\
     \times
    \exp \left\{
      -
      \sum_{\omega,p}
         \left(
           1 - \exp\left[i(\omega t + \vect{p}\cdot \vect{r})\right]
         \right)
         i \mathcal{D}^{<}_{\phi\phi}(\omega,p)
    \right\}.
  \end{multline}
\end{widetext}

This expression can be slightly simplified, since
as explained in Appendix~\ref{sec:analyt-prop-greens} below,
one has that:
\begin{equation}
  \label{eq:cancellation}
  \sum_{\omega,p}
  \left[
    i \mathcal{D}^{T}_{\phi\pi}(\omega,p) - 
    i \mathcal{D}^{\tilde{T}}_{\phi\pi}(\omega,p)
  \right] = 0.
\end{equation}
Using this result, and performing the Fourier transforms that
  appear in Eq.~(\ref{eq:two-branch-result}), one then finds the final
  form for the Green's function, as given in Eq.~(\ref{eq:final-lum}).

\section{Analytic properties of Green's functions}
\label{sec:analyt-prop-greens}

Because the use of phase and amplitude variables forces one to work in
terms of the physical Green's functions, $i \mathcal{D}^{<}$, $i
\mathcal{D}^{>}$, $i \mathcal{D}^{T}$, and $i
\mathcal{D}^{\tilde{T}}$, it is necessary to consider the analytic
properties of these Green's functions.
As discussed in Ref.~\onlinecite{Kamenev}, these Green's functions
are not independent, but for $t\ne 0$ one has
\begin{equation}
  \label{eq:tne0-relation}
  \mathcal{D}^{T} + \mathcal{D}^{\tilde{T}}
  =
  \mathcal{D}^{<} + \mathcal{D}^{>}
\end{equation}
This lack of independence is implicit in Eq.~(\ref{eq:inv-keldysh-1})
and~~(\ref{eq:inv-keldysh-2}), repeated here for convenience:
\begin{align}
  \nonumber
  {\cal D}^{<,>} &= \frac{1}{2}
  \left(
    {\cal D}^{K} 
    \mp \left[
      {\cal D}^{R} - {\cal D}^{A}
    \right]
  \right)
  \\
  {\mathcal{D}}^{T,\tilde{T}} &= \frac{1}{2}
  \left(
    {\mathcal{D}}^{K} 
    \pm \left[
      {\mathcal{D}}^{R} + {\mathcal{D}}^{A}
    \right]
  \right).
  \label{eq:rotation-summary}
\end{align}
However, at $t=0$, Eq.~(\ref{eq:tne0-relation}) does not hold.
For the case of the field-field correlations, as discussed in
Ref.~\onlinecite{Kamenev}, the correct regularisation leads to
\begin{align}
  \nonumber
  i\mathcal{D}^{T}_{\psi^{\dagger}\psi}(0)&=
  i\mathcal{D}^{\tilde{T}}_{\psi^{\dagger}\psi}(0)=
  i\mathcal{D}^{<}_{\psi^{\dagger}\psi}(0)=
  N/V
  \\
  i\mathcal{D}^{>}_{\psi^{\dagger}\psi}(0)&=
  (N+1)/V,
  \label{eq:t0-relation}
\end{align}
where $N$ is total particle number, and $V$ is volume.
The difference of form here is expected, as it encodes
important information about the equal time commutation
relations,
\begin{equation}
  \label{eq:field-commutator}
  \lim_{t\to 0}
  \left(
    i \mathcal{D}^{>}_{\psi^{\dagger}\psi}(t,r)
    -
    i \mathcal{D}^{<}_{\psi^{\dagger}\psi}(t,r)
  \right)
  = 
  \left[
    \hat{\psi}\left(\frac{r}{2}\right),
    \hat{\psi}^{\dagger}\left(-\frac{r}{2}\right)
  \right]
\end{equation}
Thus, as one expects in a path integral formulation, operator ordering
has been encoded via time ordering\cite{Kleinert}.
Written in terms of Green's functions as functions of frequency and
momentum, the left hand side of Eq.~(\ref{eq:field-commutator}) would
involve a conditionally convergent sum of terms that go like
$1/\omega$.
Preservation of commutation relations thus requires correct regularisation
of such conditionally convergent sums.
The relations for the amplitude-phase correlation functions
can be similarly found to correspond to the definition
\begin{displaymath}
  \left[
    \hat{\pi}\left(\frac{r}{2}\right),
    \hat{\phi}\left(-\frac{r}{2}\right)
  \right]
  =
  i \delta(r).
\end{displaymath}

The amplitude-phase Green's functions are found in the retarded,
advanced and Keldysh basis, but to derive the field-field correlators
we must rotate them to the forward and backward basis.
Since this includes Green's functions of non-commuting operators
evaluated at $t=0$, such as Eq.~(\ref{eq:cancellation}), it is
important to reconcile Eq.~(\ref{eq:t0-relation}) with
Eq.~(\ref{eq:rotation-summary}).
Naively, such a reconciliation does not seem possible, however the
resolution is that one must write:
\begin{align}
  i\mathcal{D}^{T}_{\psi^{\dagger}\psi}(t \to 0^+)=&
  i\mathcal{D}^{\tilde{T}}_{\psi^{\dagger}\psi}(t \to 0^-)=
  i\mathcal{D}^{<}_{\psi^{\dagger}\psi}(t=0)=  N/V
  \nonumber\\
  i\mathcal{D}^{>}_{\psi^{\dagger}\psi}(t=0)=&
  (N+1)/V
\end{align}

With such a convention, the correct regularisation of the sum in
Eq.~(\ref{eq:cancellation}) is then clear:
\begin{equation}
  \label{eq:cancellation-resolvedc}
  \left[
    i \mathcal{D}^{T}_{\phi\pi}(t \to 0^+,r=0)
    -
    i \mathcal{D}^{\tilde{T}}_{\phi\pi}(t\to 0^-,r=0)
  \right] = 0.
\end{equation}

\end{document}